\documentclass[%
reprint,
 amsmath,amssymb,
 aps,
]{revtex4-2}

\usepackage{graphicx}
\usepackage{dcolumn}
\usepackage{bm}

\usepackage{subfigure}

\begin{document}

\preprint{APS/123-QED}

\title{Calculating the Gravitational Waves Emitted from High-speed Sources}

\author{Han Yan}
\author{Xian Chen}%
 \email{Corresponding author. \\ xian.chen@pku.edu.cn}
\affiliation{%
 Department of Astronomy, School of Physics, Peking University, 100871 Beijing, China \\
and Kavli Institute for Astronomy and Astrophysics at Peking University, 100871 Beijing, China
}%
\author{Alejandro Torres-Orjuela}
\affiliation{%
 MOE Key Laboratory of TianQin Mission, TianQin Research Center for Gravitational Physics \& School of Physics and Astronomy, \\
Frontiers Science Center for TianQin, Gravitational Wave Research Center of CNSA, \\ Sun Yat-Sen University (Zhuhai Campus), Zhuhai 519082, China
}%

\date{\today}

\begin{abstract}
The possibility of forming gravitational-wave sources 
with high center-of-mass (c.m.) velocities in the
vicinity of supermassive black holes requires us to develop a method of
deriving the waveform in the observer's frame.  Here we show that in the
limit where the c.m. velocity is high but the relative
velocities of the components of the source are small, we can solve the
problem by directly integrating the relaxed Einstein field equation.  In
particular, we expand the result into multipole components which can be
conveniently calculated given the orbit of the source in the observer's
frame.  Our numerical calculations using arbitrary c.m. velocities show that
the result is consistent with the Lorentz transformation of GWs to the
leading order of the radiation field. Moreover, we show an example of using
this method to calculate the waveform of a scattering event between the
high-speed ($\sim 0.1c$) stellar objects  embedded in the accretion disk of
an active galactic nucleus.  Our multipole-expansion method not only has
advantages in analyzing the results from stellar-dynamical models but also
provides new insight into the multipole properties of the GWs emitted from a
high-speed source.  
\end{abstract}

\maketitle


\section{\label{sec:intro}Introduction }

Recent studies suggest that high-speed gravitational-wave sources could form in
the vicinity of supermassive black holes (SMBHs).  For example, it has been
shown that in the accretion disk of an active galactic nucleus (AGN),
stellar-mass binary black holes could form at a distance of tens to hundreds of
Schwarzschild radii from the central SMBH
\cite{2012MNRAS.425..460M,2019ApJ...878...85S,2020ApJ...898...25T}. The merger
rate of these binaries could be comparable to the LIGO/Virgo event rate
\cite{2018ApJ...866...66M,2019ApJ...884...22A,2020ApJ...903..133S,2020A&A...638A.119G,2020ApJ...898...25T}.
In the most extreme cases, hydrodynamic interaction with the surrounding gas
could deliver the binaries to the innermost stable circular orbit (ISCO)
\cite{2019MNRAS.485L.141C,2021MNRAS.505.1324P}. Then the center-of-mass (c.m.) velocities of the stellar-mass binaries would reach a significant fraction of the speed of light, while the relative velocities between the stellar-mass black holes are much smaller (non-relativistic).

Identifying such a high-speed source is not a trivial task. First, a constant
velocity normally induces a Doppler shift of the gravitational-wave (GW)
frequency, but the same effect can be induced by a change of the model
parameters, such as the mass \cite{2017PhRvD..95d4029B,2019MNRAS.485L.141C}, or
by including additional environmental factors in the model, such as gas
friction  (e.g.  \cite{2020ApJ...896..171C} and references therein).  Second,
if the source is accelerating, the variation of the velocity may become
detectable
\cite{2017PhRvD..95d4029B,2017ApJ...834..200M,2017PhRvD..96f3014I,2019MNRAS.488.5665W,2019PhRvD.100l4010W,2019ApJ...878...75R,2020PhRvD.101h3028T,2020PhRvD.101f3002T,2021PhRvL.126b1101Y}
but detecting it requires a space-borne GW detector since it can track the
source for sufficiently long time \cite{2019PhRvD..99b4025C}. Even in this case
there is still a parameter space where the effect is, again, degenerate with
mass \cite{2021MNRAS.502.4199X}.  For ground-based detectors, the duration of
the GW event is typically much shorter than a second, inducing the acceleration
undetectable.

Because of the difficulties in detecting a Doppler shift, several recent works
looked for other effects induced by the constant velocity but with inconclusive
results.  A well-known, standard method is to first derive the wave tensor
using the conventional quadrupole formula
\cite{1963PhRv..131..435P,1964PhRv..136.1224P} in the source frame where the
c.m. velocity is small, and then Lorentz transform it into the observer's
frame.  While the transformation induces higher modes in the radiation pattern,
whether or not the extra modes are detectable by a single detector remains
unclear \cite{2021PhRvL.127d1102T}.  An alternative method given by
Torres-Orjuela et al. is to stay in the source frame and calculate the response
of a moving detector \cite{2019PhRvD.100f3012T}.  The result agrees with the
Lorentz transformation of a single ray, indicating no detectable effect of a
constant velocity by a single detector. This result confirms the earlier
prediction based on the properties of the GWs in the weak field approximation
\cite{1983LNP...124....1T,1987thyg.book.....H}. It also confirms the
semi-quantitative analysis given by Maggiore \cite{2009GReGr..41.1667H}, in
which he considered the GWs from an elastic two-body collision system and
analyzed the aberration effect. 

These previous methods require the knowledge of the GWs in the source frame. However, the properties of the source are sometimes given not in
the rest frame of the source but in that of the observer. In particular, the
aforementioned astrophysical models for the formation of high-speed GW sources
are normally constructed in the rest frame of the central SMBH, which has
relatively small velocity with respect to the observer.  In this case, it would
be more convenient to use the physical quantities in the observer's frame to
calculate the GWs.  

Such an effort is analogous to the derivation of the
Liénard–Wiechert (L-W) potential in the electrodynamics. It not only provides
an alternative way of calculating the electromagnetic fields, but also enables
us to calculate the electromagnetic radiation from a source with both high c.m.
velocity and high c.m. acceleration. 
We notice that a derivation of  the gravitational L-W potential can be found in
the literature \cite{1999PhRvD..60l4002K}.  However, the formula (more
precisely the energy-momentum tensor) was tailored to solve the problems of the
propagation of light, gravitational lensing, or gravitomagnetism
\cite{1999PhRvD..60l4002K,2007GReGr..39.1583K} in the weak field of one or more moving
bodies (e.g. \cite{2014CQGra..31q5001Z}).  Therefore,
it is not accurate for calculating the leading order of GW, a problem not easily resolvable
in the framework of their method.

Alternatively, Press showed that the gravitational radiation of a source
extending into its own wave zone can be calculated by a method similar to
multipole expansion \cite{1977PhRvD..15..965P}. Although his formula is not
directly applicable to our problem because it diverges for a high-speed source,
it points to the importance of synthesizing different methods and including the ``effective energy-momentum
pseudo-tensor'' in the calculation to get a correct and analytical formula of GWs.   

These previous works motivate us to develop an new approach of calculating the
GWs emitted from a high-speed source directly in the observer's frame.  The
paper is organized as follows.  In Section~\ref{sec:theo}, we develop a
multipole expansion to integrate the relaxed Einstein equation which is
applicable a high-speed GW source and self-consistently includes the
contribution of the effective energy-momentum pseudo-tensor.  In
Section~\ref{sec:dis}, we compare our result with the earlier ones and show an
astrophysical example of a scattering event with a high c.m. velocity.
Finally, we discuss the significance and future extension of our work in
Section~\ref{sec:con}.  

Throughout this paper, unless otherwise indicated, we
will choose geometric units of $G = c = 1$, and the Minkowski metric is set as
$diag \left ( -1,1,1,1 \right ) $ . Latin alphabets represent three spatial
indices, and Greek alphabets represent all four indices.

\section{\label{sec:theo}Theory}

\subsection{\label{sec:level2}Multipole expansion formula}

We first derive a general multipole-expansion formula for a source with a constant c.m. velocity $\vec{\beta}$. The effect of a varying $\vec{\beta}$ will be briefly discussed in Section~\ref{sec:con}.

As we mentioned in the Introduction, the quadrupole formula is derived in the rest frame of the source and in the limit of low velocity, and hence not suitable for our problem. Therefore,
we start from the ``relaxed Einstein field equations'' which is a result of the Landau-Lifshitz formulation \cite{1975ctf..book.....L} and valid for high velocities. We integrate it to get
\begin{eqnarray}
h^{\mu \nu}(t, \vec{x})=4 \int \frac{\tau^{\mu \nu}\left(t-\left|\vec{x}-\vec{x}^{\prime}\right|, \vec{x}^{\prime}\right)}{\left|\vec{x}-\vec{x}^{\prime}\right|} \mathrm{d}^{3} x^{\prime}\label{eq:hmn}
\end{eqnarray}
(see \cite{2014grav.book.....P} for details). Note that $h^{\mu \nu}$ given by
the above equation is an exact solution to the relaxed Einstein field equations, and its linear order
corresponds to the GW in the ``weak field approximation''.

Here $\tau^{\mu \nu}$ is the ``effective energy-momentum pseudo-tensor'' contributed by both mass and
the gravitational field \cite{2014grav.book.....P}. To simplify the following derivation, we rewrite Equation~(\ref{eq:hmn}) as:
\begin{equation}
    \psi(t, \vec{x})=\int \frac{\mu\left(t-\left|\vec{x}-\vec{x}^{\prime}\right|, \vec{x}^{\prime}\right)}{\left|\vec{x}-\vec{x}^{\prime}\right|} \mathrm{d}^{3} x^{\prime} ~.\tag{1$'$}\label{eq:1p}
\end{equation}

This integral can be computed with a multipole expansion. First, we
modify the integrand in Eq.~(\ref{eq:1p}) using the Dirac $\delta \text{-}$ function,
\begin{eqnarray}
\frac{\mu\left(t-\left|\vec{x}-\vec{x}^{\prime}\right|, \vec{x}^{\prime}\right)}{\left|\vec{x}-\vec{x}^{\prime}\right|} &=&\int \frac{\mu\left(t-\left|\vec{x}-\vec{x}^{\prime}\right|, \vec{y}\right)}{\left|\vec{x}-\vec{x}^{\prime}\right|} \delta^{3}\left(\vec{y}-\vec{x}^{\prime}\right) \mathrm{d}^{3} y \nonumber \\
&\equiv& \int g\left(t, \vec{x}, \vec{x}^{\prime}, \vec{y}\right) \delta^{3}\left(\vec{y}-\vec{x}^{\prime}\right) \mathrm{d}^{3} y ~. \label{eq:2}
\end{eqnarray}
Second, noticing that the c.m. velocity, $\vec{\beta}$, is much greater than the relative velocity between the two components of the binary, we introduce an important variable $\vec{\Delta}\equiv \vec{x}^{\prime}-\vec{\beta}\left(t-\left|\vec{x}-\vec{x}^{\prime}\right|\right)$ to extract the small displacement relative to the center of mass.
The new set of variables $(t,\vec{x},\vec{\Delta},\vec{y})$ replaces the old set $(t,\vec{x},\vec{x}^{\prime},\vec{y})$. 
This replacement allows us to do the Maclaurin series expansion of the function $g(t, \vec{x}, \vec{\Delta}, \vec{y})$, in Eq.~(\ref{eq:2}) with respect to $\vec{\Delta}$. 
Third, after the expansion we restore the variable $\vec{x}^{\prime}$ using the relationship
\begin{eqnarray}
\frac{\partial g(t, \vec{x}, \vec{\Delta}, \vec{y})}{\partial \Delta^{i}}=-
\frac{\partial g(t, \vec{x}, \vec{\Delta}, \vec{y})}{\partial x^{i}}
~, \label{eq:3}
\end{eqnarray}
similar to the operation in Ref.~\cite{2014grav.book.....P}. 

Finally, Eq.~(\ref{eq:1p}) becomes a multipole-expansion formula
\begin{eqnarray}
\psi(t, \vec{x})=&& \sum_{l=0}^{\infty}  \frac{(-1)^{l}}{l !}  \nonumber  \int  \mathrm{d}^{3} x^{\prime} \\ &&  \vec{\Delta}^{L}  \partial_{L} \frac{\mu\left(t-|\vec{x}-\vec{\beta} \tau|, \vec{x}^{\prime}\right)}{|\vec{x}-\vec{\beta} \tau|} ~. \label{eq:4}
\end{eqnarray}
Here  $\tau$ is the c.m. retarded time which satisfies $\tau=t-|\vec{x}-\vec{\beta} \tau|$, and  ${L}$ always denotes $l$ different space indices (e.g.: $L=ijk$ for $l=3$, see Ref. \cite{2014grav.book.....P} for details). 

Now we will simplify Eq.~(\ref{eq:4}) to get a more familiar form containing the time derivatives of the multipole moments of the source. More specifically, we replace the partial derivatives $\partial_{L}$ in 
Eq.~(\ref{eq:4}) by the terms proportional to 
$n_{L}\left({\mathrm{d}}/{\mathrm{d} \tau}\right)^{l}$, where $\hat{n} \equiv \vec{x} / R$ is the wave vector and 
$R\equiv \left | \vec{x} \right | $.
Keeping the terms to the radiation order $O\left ( {1}/{R}  \right ) $,
we have
\begin{eqnarray}
\psi(t, \vec{x})= && \frac{1}{R} \sum_{l=0}^{\infty} \frac{1}{l !}\left(\frac{1}{1-\hat{n} \cdot \vec{\beta}}\right)^{l} \nonumber \\ && \int \mathrm{d}^{3} x^{\prime}\left(\hat{n} \cdot\vec{\Delta}\right)^{l}\left(\frac{\mathrm{d}}{\mathrm{d} \tau}\right)^{l} \mu\left(\tau, \vec{x}^{\prime}\right) ~.\label{eq:5}
\end{eqnarray}

To move the time derivatives outside the integral, we do the integration by parts. 
Notice that now the function $\left(\hat{n} \cdot\vec{\Delta}\right)^{l}$
in the integrand depends on $\tau$. Therefore, we can not directly take the time derivatives outside the
integral as we normally do in the c.m. frame. Nevertheless, after some algebra we get 
\begin{eqnarray}
&& \psi(t, \vec{x}) = \frac{1}{R}   \frac{1}{1-\hat{n} \cdot \vec{\beta}} \int \mathrm{d}^{3} x^{\prime} \mu \nonumber \\
&& +\frac{1}{R}\frac{1}{(1-\hat{n} \cdot \vec{\beta})^{3}} \frac{\mathrm{d}}{\mathrm{d} \tau} \int \mathrm{d}^{3} x^{\prime} \hat{n} \cdot\vec{\Delta}  \mu \nonumber \\
&& + \frac{1}{2R}\frac{1}{(1-\hat{n} \cdot \vec{\beta})^{5}} \nonumber \\ &&\times \left(\frac{\mathrm{d}}{\mathrm{d} \tau}\right)^{2} \int \mathrm{d}^{3} x^{\prime}\left(\hat{n} \cdot\vec{\Delta}\right)^{2}  \mu   ~.\label{eq:6}
\end{eqnarray}
The details of the derivation can be found in Appendix A.
Notice that Eq.~(\ref{eq:6}) contains only the leading (quadrupole) order of GW. We will discuss the possibility
of including higher-order terms at the end of this paper. 

\subsection{\label{sec:level2}Point-mass sources}

In our problem the source is composed of compact objects, which can be approximated by point masses.
We refer to the mass and position of the $m$th component as, respectively,  $M_m$ and $\vec{r}_{m}$.
The corresponding Lorentz factor is $\gamma_m$.
In this case, Eq.~(\ref{eq:6}) leads to
\begin{eqnarray}
&& h^{i j}(t, \vec{x}) =\frac{1}{1-\hat{n} \cdot \vec{\beta}} \frac{1}{R} Q^{0}(t, \vec{x}, \tau(t, \vec{x})) \nonumber \\
&&+\frac{1}{(1-\hat{n} \cdot \vec{\beta})^{2}} \frac{1}{R} \frac{\mathrm{d}}{\mathrm{d} \tau}\left(n_{k} Q^{k}(\tau)\right) \nonumber \\
&&+\frac{1}{2} \frac{1}{(1-\hat{n} \cdot \vec{\beta})^{3}} \frac{1}{R}\left(\frac{\mathrm{d}}{\mathrm{d} \tau}\right)^{2}\left(n_{k l} Q^{k l}(\tau)\right) ~, \label{eq:7}
\end{eqnarray}
where we have defined three ``mass multipole moments'',
\begin{eqnarray}
Q^{0} &&=4 \int \tau^{i j} \mathrm{d}^{3} x^{\prime} \nonumber \\
n_{k} Q^{k} &&=4 \sum_{m} \gamma_{m} M_{m} v_{m}^{i} v_{m}^{j} \hat{n} \cdot \vec{\delta}_{m} \nonumber \\
n_{k l} Q^{k l} &&=4 \sum_{m} \gamma_{m} M_{m} v_{m}^{i} v_{m}^{j}\left(\hat{n} \cdot \vec{\delta}_{m}\right)^{2} ~,\label{eq:8}
\end{eqnarray}
and a new variable $\vec{\delta}_{m}(\tau) \equiv \vec{r}_{m}(\tau)-\vec{\beta} \tau$.
Note that in Eq.~(\ref{eq:7}), we have neglected the octupole and higher-order terms.

The reason we kept $\tau^{ij}$ in the calculation of $Q^{0}$ is that both mass and the gravitational ``field energy'' contribute to this term. To express $Q^{0}$ only in terms of the mass, we recall that normally 
in the c.m. frame we use the conservation law of the energy-momentum tensor to get a second-order time-differentiation form. We do a similar calculation here but take into account the c.m. velocity $\vec{\beta}$ (see details in the Appendix B), which results in
\begin{equation}
\tau^{i j}\left(\tau, \vec{x}^{\prime}\right) =\frac{1}{2} \partial_{00}\left(\tau^{00}\left(x^{\prime i}-\beta^{i} \tau\right)\left(x^{\prime j}-\beta^{j} \tau\right)\right)+\partial_{k}(\cdot) ~.\label{eq:9}
\end{equation}
Here $\tau^{00}$ is predominantly contributed by the mass. The term $\partial_{k}(\cdot)$ represents the non-radiative total-differentiation part, which can be discarded in the calculation of GWs.  When $\vec{\beta} = 0$, Eq.~(\ref{eq:9}) recovers the standard result in the c.m. frame. 

Now the only part that has not been calculated in Eq.~(\ref{eq:7}), i.e., $Q^{0}$, can be derived by integrating  
Eq.~(\ref{eq:9}). The final result for $h^{ij}$ is
\begin{eqnarray}
&& h^{i j}(t, \vec{x}) =\frac{2}{1-\hat{n} \cdot \vec{\beta}} \frac{1}{R}\left(\frac{\mathrm{d}}{\mathrm{d} \tau}\right)^{2}\left[\sum_{m} \gamma M \delta^{i} \delta^{j}\right] \nonumber\\
&&+\frac{4}{(1-\hat{n} \cdot \vec{\beta})^{2}} \frac{1}{R} \frac{\mathrm{d}}{\mathrm{d} \tau}\left[\sum_{m} \gamma M v^{i} v^{j} \hat{n} \cdot \vec{\delta}\right] \nonumber\\
&&+\frac{2}{(1-\hat{n} \cdot \vec{\beta})^{3}} \frac{1}{R}\left(\frac{\mathrm{d}}{\mathrm{d} \tau}\right)^{2}\left[\sum_{m} \gamma M v^{i} v^{j}\left(\hat{n} \cdot \vec{\delta}\right)^{2}\right] ~,\label{eq:10}
\end{eqnarray}
where $\vec{\delta}_{m}(\tau) \equiv \vec{r}_{m}(\tau)-\vec{\beta} \tau$, and the index $m$ on the right of $\sum_{m}$ are omitted for simplicity.
We will discuss the meaning of the above equation in the next section.

To see more clearly the dependence on the c.m. velocity $\beta$ in the low-velocity limit, 
we expand Eq.~(\ref{eq:10}) to different orders of $\beta$. Keeping the zeroth and first order terms, we get
\begin{equation}
\begin{aligned}
h^{i j}(t, \vec{x}) & =\frac{2}{1-\hat{n} \cdot \vec{\beta}} \frac{1}{R}\left(\frac{\mathrm{d}}{\mathrm{d} \tau}\right)^{2}\left[\sum_{m} M \delta^{i} \delta^{j}\right] \nonumber \\
&+\frac{4}{1-2 \hat{n} \cdot \vec{\beta}} \frac{1}{R} \frac{\mathrm{d}}{\mathrm{d} \tau}\left[\sum_{m} M v^{i} v^{j} \hat{n} \cdot \vec{\delta}\right] ~.
\end{aligned}
\tag{10$'$}\label{eq:10p}
\end{equation}
We see that when $\beta=0$ (zeroth order), the first term in the last equation recovers the standard  
quadrupole moment formula in the c.m. frame. The second term contains only part of the octupole moment
because of the approximation we did since Eq.~(\ref{eq:6}).

As a reminder, we should mention here that the second line of Eq.~(\ref{eq:10}) or Eq.~(\ref{eq:10p}) does seem to have a dipole term at the leading order. We notice that $v^i$ contains a $\beta ^i$ and therefore the apparent leading order term is the time differentiation of $\beta ^{i} \beta ^{j} \hat{n} \cdot \sum_{m} \gamma M  \vec{\delta}$. However, this kind of terms will cancel out and leave a quadrupole term simply because of the conservation of the total momentum, as our numerical calculations  in Section~\ref{sec:dis} will prove. People should then not be misled and carefully handle those
apparent ``dipole moment terms'' induced by c.m. velocity when calculate the wave templates of a high-speed source.

\section{\label{sec:dis}Test and application}

\subsection{Compare with Lorentz transformation}

As we have mentioned in the Introduction, it is commonly accepted that one can use Lorentz transformation to derive the GWs from a moving source in the observer's frame.
Therefore, we compare our result with the result derived from Lorentz transformation.

We consider a simple but instructive example in which the source is an one-dimensional harmonic oscillator with a 
single frequency (in its c.m. frame). We assign an arbitrary source velocity and
calculate its trajectory in the observer's frame using a Lorentz
transformation and numerical interpolation. We then compute the GW in the observer's frame and compare it
with the result given by Eq.~(\ref{eq:10}). We find that the results are the same at the leading (quadrupole-moment) order \footnote{The code used for this calculation can be found at https://github.com/StrelitziaHY/GW-transformation}. The agreement suggests that our method described in Section~\ref{sec:theo}
is feasible.

It is worth mentioning that our formula not only calculates the amplitude of the GW in the observer's frame, but also self-consistently produces the relativistic Doppler shift of the frequency.
More specifically, the 
the Lorentz factor comes from the time dilation effect and
the geometrical factor $(1-\hat{n} \cdot \vec{\beta})$ comes from the retardation effect (e.g., see Eq.~(\ref{eq:5})).
The two factors combined give the correct Doppler factor.

\subsection{\label{sec:compare}Compare with previous work}

Torres-Orjuela et al. presented another way of calculating the GW signal \cite{2019PhRvD.100f3012T}. They stayed in the rest frame of the source and studied the response of a fast-moving detector. Now we compare their result with ours and show that they are also consistent.

We consider a more realistic source which is an equal-mass binary moving at a relativistic c.m. velocity. In the following, the initial conditions are all specified in the rest frame of the source, to allow easier comparison with the results of of Torres-Orjuela et al.
We assume that the binary has a total mass of $M = 2$ with arbitrary unit and the orbital velocity is $v_{0} = 0.01$. The c.m. coincides with the origin of the coordinates and the orbital plane is aligned with the  $x-y$ plane. Initially, the observer is on the $z$ axis at a distance of $z = 10^{10}$ from the origin. The velocity of the observer is $(-0.6, 0, 0)$, i.e., it is anti-parallel with the $x$ axis. 
Having defined the initial conditions, we calculate the orbit of the binary in the observer's frame 
using Lorentz transformation
and derive GW amplitude $h^{ij}$ using our Eq.~(\ref{eq:10}). 

To calculate the response of a detector, we have to specify the antenna patterns. Therefore, we
follow Torres-Orjuela et al. and consider a detector with two orthogonal arms with equal length. In the source frame, the two arms are pointing in the directions $\hat{n}_{a}=(0,1,0)$ and $\hat{n}_{b}=(\cos \theta, 0, \sin \theta)$. Given such a detector, we first Lorentz transform the directions of the arms and the wave vector into the
observer's frame to get $\hat{n}_{a}^{\prime}$, $\hat{n}_{b}^{\prime}$, and $\hat{n}^{\prime}$, and then compute the response with
\begin{equation}
    F\left(\hat{n}_{a}^{\prime}, \hat{n}_{b}^{\prime}\right)=\Lambda_{i j, k l}(\hat{n}^{\prime}) h^{k l}\left(\hat{n}_{a}^{i\prime} \hat{n}_{a}^{j\prime}-\hat{n}_{b}^{i\prime} \hat{n}_{b}^{j\prime}\right) ~,\label{eq:11}
\end{equation}
where $\Lambda_{i j, k l}(\hat{n}')$ is the standard transverse-traceless (TT) projection operator (see its definition in  Ref.~\cite{2009GReGr..41.1667H}).

\begin{figure}
\includegraphics[width=1\linewidth]{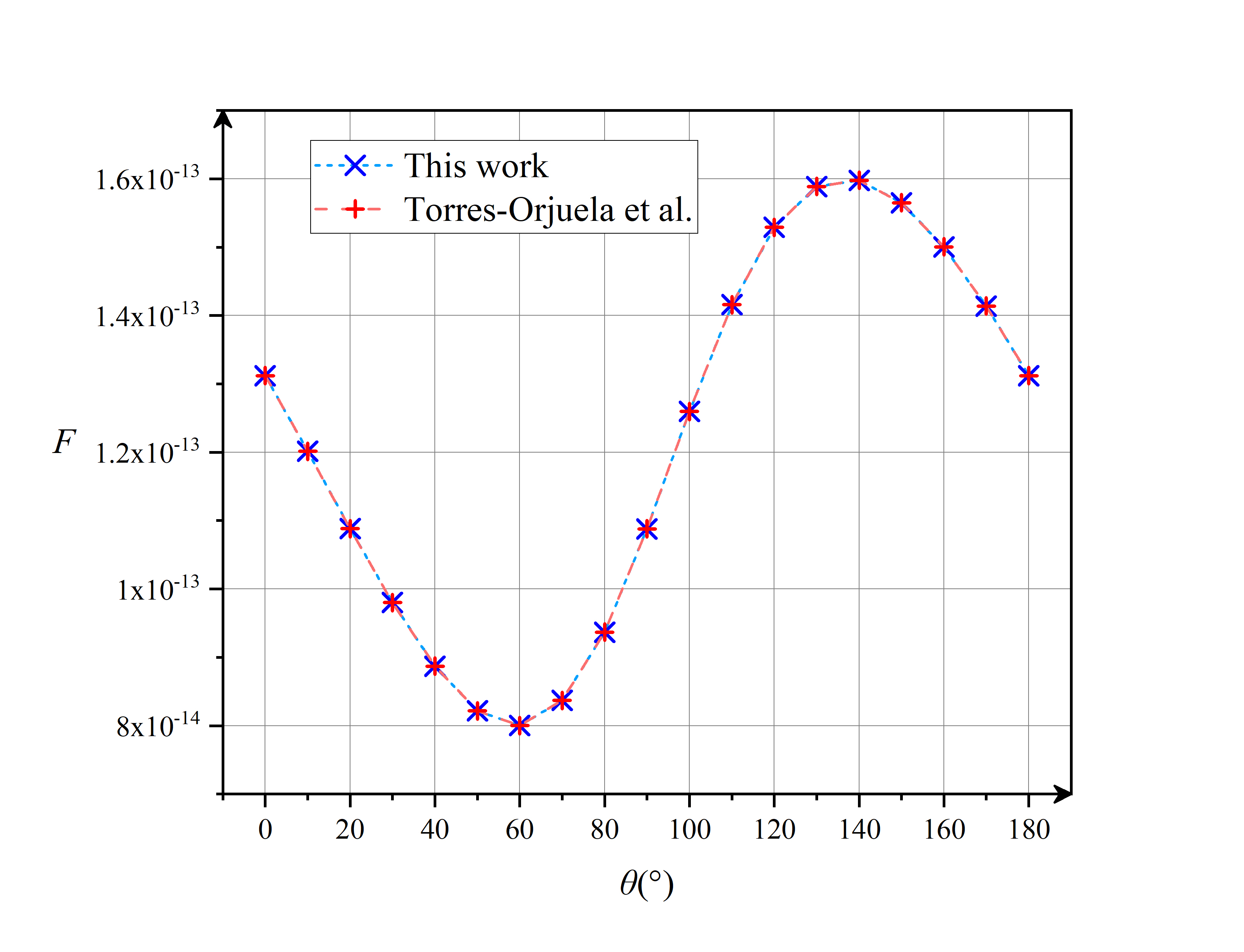}
\caption{\label{fig:1}Detector response to a moving binary source as a function of the orientation of the arms.
The two symbols refer to the results calculated using two different methods. They agree within a relative
error of $4.3\times 10^{-6}$.
}
\end{figure}

Now we can compare the result from Eq.~(\ref{eq:11}) with that from Eq.~(29) in the work of Torres-Orjuela et al. \cite{2019PhRvD.100f3012T}. Fig.~\ref{fig:1} shows that over the entire range of $\theta\in[0^{\circ}, 180^{\circ}]$, the two results agree well, with a small relative error of about $4.3\times 10^{-6}$
when $\theta$ varies.

\subsection{\label{sec:level2}Scattering between stellar objects in AGN disk}

Our work is partly motivated by the high-velocity scatters between stellar objects which may happen in the accretion disk surrounding a SMBH. Now we calculate the GW emitted during such a scatter. 

We consider a toy model with two equal-mass stellar objects ($m_1=m_2=m$), both on corotating Keplerian circular orbits around a SMBH of mass $M$. To simplify the problem, we start our analysis when the two objects are already close, separated by a tangential distance of $d$ and a radial distance of $r$. As long as $d$ and $r$ are sufficiently smaller than the distance to the central SMBH, $D$, we can approximate the c.m. velocity of these two objects with $\beta = \sqrt{M/D} $, as well as neglect their earlier mutual interaction. These initial conditions result in a difference in the initial tangential velocities of the two objects, which can be calculated with $v_c \simeq  r\beta/2D$. 

The scattering process can be further simplified if
the duration of the scattering ($\simeq \sqrt{\left ( d^{2} + r^{2}   \right ) ^{3/2}/m   }$) is much shorter than the orbital period ($\simeq D/\beta$) around the SMBH, in which case we can 
adopt an impulsive approximation and neglect the tidal force of the SMBH during the interaction.
The validity of this approximation requires that
\begin{equation}
    \left ( \frac{d}{r} \right ) ^{2} +1  \ll \left ( \frac{\beta }{2v_c}  \right )^{2}   \left ( \frac{m}{M}  \right )^{\frac{2}{3}}  ~. \label{eq:12}
\end{equation}

The values of the parameters are chosen as follows.  We set ${m}/{M}  = 10^{-4}
$, $\beta = 0.1$, and $v_c = 10^{-3}$. Consequently, we have $r = 2\times
10^{4} m $, indicating that initially the two stellar objects are far apart
from merger. Given these parameters, Eq.~(\ref{eq:12}) becomes $\left ( d/r
\right ) ^{2} +1 \ll 5.4$. Therefore, we choose $d = 0.5 r$ to comply with this
requirement. The trajectories of the two objects are first obtained in their
c.m. frame and then Lorentz transformed into the observer's frame, i.e. the
rest frame of the SMBH. The coordinates in the observer's frame are chosen
mostly in the same way as in Section~\ref{sec:compare}. In particular, we
choose the same orientation (relative to the coordinate axes) for the  orbit
plane and the same direction for the c.m. velocity.  The wave vector is set to
$\hat{n} = \left ( 0.5,0,\sqrt{1-0.5^{2} }  \right )$ as an example, and the
distance to the observer is denoted as $R$ as before. Both $\hat{n}$ and $R$
are defined in the observer's frame. 

\begin{figure}
\includegraphics[width=1\linewidth]{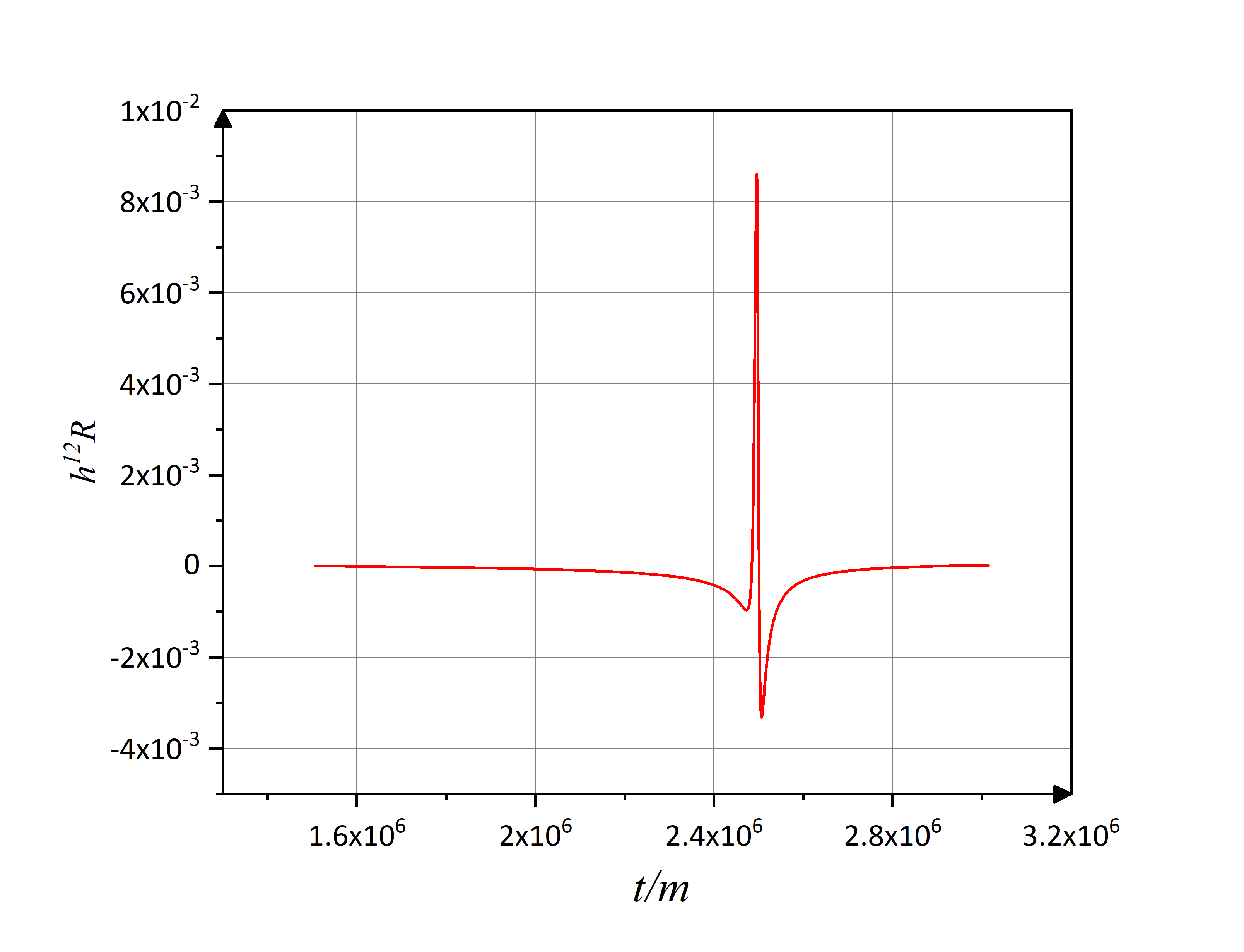}
\caption{\label{fig:2}Evolution of the $h^{12}$ component of the GWs emitted from a scattering event happening close to a SMBH. Note that the result is normalized by the distance factor $1/R$.}
\end{figure}

Using Eq.~(\ref{eq:10}) again, we calculate the $(1,2)$ component of the GWs,
$h^{12}$, and show the evolution of $h^{12}R$ in Fig.~\ref{fig:2}. Note that
this component is different from the standard cross polarization because we did
not performed a TT projection yet.  The sharp spike at $t\simeq2.5\times 10^{6}
m$ corresponds to the GW radiated during the pericenter passage.  It is
important to understand that the calculation of the waveform only requires the
knowledge of the trajectories in the SMBH frame. Therefore, our method will
have advantages in analyzing the scattering events from numerical N-body
simulations, which are normally performed in the SMBH frame
\cite{1982Ap&SS..85..231D,2009MNRAS.395.2127O,2017PhRvD..96h4009B,2018MNRAS.474.5672L}.    

\section{\label{sec:con}Discussion and Conclusion}

In this paper, we have developed a new method of calculating the GW radiation
of a high-speed source. In particular, we expanded the relaxed Einstein
equation with a special form of multipole expansion in the observer's frame.  At the
leading (quadrupole) order, our result recovers the previous one derived from
Lorentz transformation, as we have shown in Section~\ref{sec:dis}. 

One advantage of our method is that it directly calculates the GWs radiated by
a source with a high c.m. velocity relative to the observer.  The calculation,
using Eq.~(\ref{eq:10}), only requires knowledge of the orbit of the source in
the observer's frame.  With an additional TT projection, one can conveniently
get the waveform in the observer's frame.  The conventional Lorentz
transformation, however, first requires a transformation of the orbit into the
source frame, then uses the standard quadrupole formula to calculate the
waveform in the source frame, and finally performs an inverse Lorentz
transformation to get the observed waveform.  For this reason, our method is
useful in analyzing the results of N-body numerical simulations, where the
orbits of the GW sources are normally given in the observer's fame.

Our work also generalizes the method of using multipole expansion to calculate
GWs. While the conventional multipole-expansion formula is derived in the rest
frame of the source, our method relaxes the requirement on the c.m. velocity so
that it is applicable to a source with arbitrary c.m. velocity.  As far as we
know, this is the first work where the form of such a generalized expansion is
explicitly spelt out.

One can see our result as 
an analog to the L-W potential in electrodynamics. It is used to calculate the electromagnetic fields generated by an arbitrarily-moving source. In particular,
such a potential is more useful than the conventional Lorentz transformation
when the velocity of the source is not constant. However, there is a key
difference between our method and the derivation of the L-W potential in electrodynamics. 
In our problem, there is a ``pseudo field energy-momentum tensor'' in $\tau^{\mu \nu}$, which is absent in the standard electrodynamics.  
For this reason, we cannot use the equation
\begin{eqnarray}
h^{\mu \nu } \sim  \frac{U ^{\mu } U ^{ \nu } \left ( t^{\ast } , \vec{x}^{\ast }   \right ) }{R^{\ast } - \hat{n} \cdot \vec{R}^{\ast } }, 
\end{eqnarray}
like the one normally used in electrodynamics, to calculate the GWs because it
is missing the part contributed by the field.  Here ``$*$'' means retardation
and $U ^{\mu }$ is the 4-velocity of the source.  Our multipole expansion has
been carefully derived to properly treat this issue.  We note that our
treatment of the aforementioned difference is not new.  In fact, in the
derivation of the conventional quadrupole formula for GWs, one has to take a
step similar to Eq.~(\ref{eq:9}) but with $\vec{\beta} = 0$ to correctly
account for the contribution from the field. Otherwise, a direct integration
of the weak-field equations will give a spurious result.

Finally, we point out two directions for future work which could broaden the
applications of our result.  First, one can consider a varying c.m. velocity,
$\vec{\beta }(\tau) $. The acceleration could possibly give rise to a series of
higher-order multipole moments and hence induce discernible signature in the
waveform. To include the effects of acceleration, Eq.~(\ref{eq:3}) is still
tenable but the analysis from Eq.~(\ref{eq:5}) has to be revised, because those
derivatives $\left({\mathrm{d}}/{\mathrm{d}} \tau\right)^{l}$ inside the
integration will also act on $\vec{\beta }$.  Second, one can also consider the
higher-order modes of GW radiation, which will be particularly important for
binaries with high orbital eccentricities.  In this case, our derivation until
Eq.~(\ref{eq:5}) will be correct. However, the following  integration by parts
needs to be revised to keep more terms, which we have omitted in the current
paper.  In addition, one has to re-assess the relative importance of the terms
coming from the ``pseudo field energy-momentum tensor'', since Eq.~(\ref{eq:9})
can only get around this problem (which means we don't need to actually
calculate these field energy-momentum tensors) in quadrupole order.

\begin{acknowledgments}
This work is supported by the National Key Research and Development Program of
	China Grant No. 2021YFC2203002 and the National Science Foundation of
	China grants No. 11991053. ATO acknowledges support from the Guangdong Major Project of Basic and Applied Basic Research (Grant No. 2019B030302001) and the China Postdoctoral Science Foundation (Grant No. 2022M723676). The authors would like to thank Rui Xu and Yun Fang for many helpful discussions.
\end{acknowledgments}

\appendix

\section{Detailed derivations of Eq.~(\ref{eq:6})}

The first step to get Eq.~(\ref{eq:6}) is to investigate the result of
$\left(\frac{\mathrm{d}}{\mathrm{d} \tau}\right)^{l}$ acting on $\left(\hat{n}
\cdot\vec{\Delta}\right)^{l}$, so that 
we can handle the integration in Eq.~(\ref{eq:5}) by parts. This step
is performed according to the definitions of $\tau$ at
the proper order,
\begin{eqnarray}
&&\left(\frac{\mathrm{d}}{\mathrm{d} \tau}\right)^{k}\left(\hat{n} \cdot\vec{\Delta}\right)^{l} \nonumber\\
=&&\left(\frac{\mathrm{d}}{\mathrm{d} \tau}\right)^{k}\left[\hat{n} \cdot\left(\vec{x}^{\prime}-\vec{\beta} \left(t-\left|\vec{x}-\vec{x}^{\prime}\right|\right)\right)\right]^{l} \nonumber\\
=&&\left(\frac{\mathrm{d}}{\mathrm{d} \tau}\right)^{k}\left(\hat{n} \cdot\left(\vec{x}^{\prime}-\vec{\beta} \hat{n} \cdot \vec{x}^{\prime}-\vec{\beta}(1-\hat{n} \cdot \vec{\beta}) \tau\right)\right)^{l} \nonumber\\
=&& \frac{l !}{(l-k) !}(-\hat{n} \cdot \vec{\beta}(1-\hat{n} \cdot \vec{\beta}))^{k} \nonumber \\
&& \times\left(\hat{n} \cdot\vec{\Delta}\right)^{l-k} ~.
\end{eqnarray}

The result of the integration by parts contains several coefficients, such as
\begin{eqnarray}
&& \int A \times \mathrm{d}^{l} B \nonumber\\
=&& c_{l} \times \mathrm{d}^{l} \int A B \nonumber\\
+&& c_{0} \times \int \mathrm{d}^{l} A \times B \nonumber\\
+&& c_{1} \times \mathrm{d}^{1} \int \mathrm{d}^{l-1} A \times B \nonumber\\
+&& c_{2} \times \mathrm{d}^{2} \int \mathrm{d}^{l-2} A \times B \nonumber\\
+&& \ldots. \label{eq:A2}
\end{eqnarray}
After some proper reordering and arrangement, we find that
the coefficients in Eq.~(\ref{eq:A2}) are
\begin{eqnarray}
&& c_{l}=1 \nonumber\\
&& c_{0}=(-1)^{l} \quad, \quad l \geq 1 \nonumber\\
&& c_{1}=(-1)^{l-1} \times l \quad, \quad l \geq 2 \nonumber\\
&& c_{2}=(-1)^{l-2} \times \frac{l(l-1)}{2} \quad, \quad l \geq 3 ~.
\end{eqnarray}

Based on the above properties, Eq.~(\ref{eq:5}) can be transformed into the following form,
\begin{eqnarray}
\psi(t, \vec{x}) &&=\sum_{l=0}^{\infty} \frac{1}{l !} \frac{1}{R} \frac{1}{(1-\hat{n} \cdot \vec{\beta})^{l}}\left(\frac{\mathrm{d}}{\mathrm{d} \tau}\right)^{l} \nonumber
\\&& \times \int \mathrm{d}^{3} x^{\prime}\left(\hat{n} \cdot\vec{\Delta}\right)^{l} \times \mu \nonumber\\
&&+\sum_{l=1}^{\infty} \frac{l}{R}(\hat{n} \cdot \vec{\beta})^{l} \int \mathrm{d}^{3} x^{\prime} \mu \nonumber \\
&&+\sum_{l=2}^{\infty} \frac{l}{R} \frac{(\hat{n} \cdot \vec{\beta})^{l-1}}{1-\hat{n} \cdot \vec{\beta}} \frac{\mathrm{d}}{\mathrm{d} \tau} \int \mathrm{d}^{3} x^{\prime} \hat{n} \cdot\vec{\Delta} \times \mu \nonumber\\
&&+\sum_{l=3}^{\infty} \frac{l(l-1)}{4 R} \frac{(\hat{n} \cdot \vec{\beta})^{l-2}}{(1-\hat{n} \cdot \vec{\beta})^{2}}\left(\frac{\mathrm{d}}{\mathrm{d} \tau}\right)^{2} 
\nonumber\\&& \times \int \mathrm{d}^{3} x^{\prime}\left(\hat{n} \cdot\vec{\Delta}\right)^{2} \times \mu ~. \label{eq:A4}
\end{eqnarray}

We then calculate several series summations about $l$ in Eq.~(\ref{eq:A4}), except in the first term. Define the series as $S(x, m)\equiv\sum_{l=1}^{\infty} l^{m} x^{l} \quad (m \in N, |x|<1 )$, and we simply write down the results we need,
\begin{eqnarray}
&& S(x, 0)=\frac{x}{1-x} \nonumber\\
&& S(x, 1)=\frac{x}{(1-x)^{2}} \nonumber\\
&& S(x, 2)=\frac{x(x+1)}{(1-x)^{3}} ~.
\end{eqnarray}

The results of the summations in Eq.~(\ref{eq:A4}) are then calculated as
\begin{eqnarray}
\psi(t, \vec{x})=&& \sum_{l=0}^{\infty} \frac{1}{l !} \frac{1}{R} \frac{1}{(1-\hat{n} \cdot \vec{\beta})^{l}} \left(\frac{\mathrm{d}}{\mathrm{d} \tau}\right)^{l}\nonumber \\ && \times  \int \mathrm{d}^{3} x^{\prime}\left(\hat{n} \cdot\vec{\Delta}\right)^{l} \times \mu \nonumber\\
&&+\frac{1}{R} \frac{\hat{n} \cdot \vec{\beta}}{1-\hat{n} \cdot \vec{\beta}} \int \mathrm{d}^{3} x^{\prime} \mu \nonumber\\
&&+\frac{1}{R} \frac{\hat{n} \cdot \vec{\beta}(2-\hat{n} \cdot \vec{\beta})}{(1-\hat{n} \cdot \vec{\beta})^{3}} \nonumber \\ 
&& \times\frac{\mathrm{d}}{\mathrm{d} \tau} \int \mathrm{d}^{3} x^{\prime} \hat{n} \cdot\vec{\Delta} \times \mu \nonumber\\
&&+\frac{1}{4 R} \frac{2 \hat{n} \cdot \vec{\beta}\left((\hat{n} \cdot \vec{\beta})^{2}-3 \hat{n} \cdot \vec{\beta}+3\right)}{(1-\hat{n} \cdot \vec{\beta})^{5}} \left(\frac{\mathrm{d}}{\mathrm{d} \tau}\right)^{2} \nonumber \\ && \times  \int \mathrm{d}^{3} x^{\prime}\left(\hat{n} \cdot\vec{\Delta}\right)^{2}  \times \mu ~. \label{eq:A6}
\end{eqnarray}

Eliminate $l > 2$ terms (i.e., higher modes of GWs) in the first line and merger the similar terms, we will get  Eq.~(\ref{eq:6}).

\section{Detailed derivations of Eq.~(\ref{eq:9})} 

The conventional result deduced from the conservation law in the c.m. frame (the first line of Eq.~(\ref{eq:B1})) is still workable, because this deduction only requires the conservation law, without any assumption about the reference frame. Based on this result, we can transform it to deduct the contributions of c.m. movement as 
\begin{eqnarray}
\tau^{i j}\left(\tau, \vec{x}^{\prime}\right) &&=\frac{1}{2} \partial_{00}\left(\tau^{00} x^{\prime i} x^{\prime j}\right)+\partial_{k}(\cdot) \nonumber\\
&&=\frac{1}{2} \partial_{00}\left(\tau^{00}\left(x^{\prime i}-\beta^{i} \tau+\beta^{i} \tau\right)\left(x^{\prime j}-\beta^{j} \tau+\beta^{j} \tau\right)\right) \nonumber\\
&&=\frac{1}{2} \partial_{00}\left(\tau^{00}\left(x^{\prime i}-\beta^{i} \tau\right)\left(x^{\prime j}-\beta^{j} \tau\right)\right) \nonumber\\
&&+\frac{1}{2} \partial_{00} \tau^{00} \left(\left(x^{\prime i}-\beta^{i} \tau\right) \beta^{j} \tau+\left(x^{\prime j}-\beta^{j} \tau\right) \beta^{i} \tau\right)
\nonumber\\
&&+\frac{1}{2} \partial_{00} \left( \tau^{00} \beta^{i} \beta^{j} \tau^{2}\right) ~. \label{eq:B1}
\end{eqnarray}

The first order time derivatives in the last two terms can be calculated as
\begin{eqnarray}
&&\partial_{0} \tau^{00} \left(\left(x^{\prime i}-\beta^{i} \tau\right) \beta^{j} \tau+\left(x^{\prime j}-\beta^{j} \tau\right) \beta^{i} \tau\right)
\nonumber\\
&&+ \partial_{0} \left( \tau^{00} \beta^{i} \beta^{j} \tau^{2}\right) \nonumber\\
&&=-\left(\partial_{k} \tau^{0 k}\right)\left(\left(x^{\prime i}-\beta^{i} \tau\right) \beta^{j} \tau+\left(x^{\prime j}-\beta^{j} \tau\right) \beta^{i} \tau\right) \nonumber\\
&& -\left(\partial_{k} \tau^{0 k}\right)\left(\beta^{i} \beta^{j} \tau^{2}\right) 
\nonumber \\
&&+\tau^{00}\left(x^{\prime i} \beta^{j}+x^{\prime j} \beta^{i}-2 \beta^{i} \beta^{j} \tau\right) \nonumber\\
&&=\tau^{0 k}\left(\delta_{k}^{i} \beta^{j} \tau+\delta_{k}^{j} \beta^{i} \tau\right)+\tau^{00}\left(x^{\prime i} \beta^{j}+x^{\prime j} \beta^{i}-2 \beta^{i} \beta^{j} \tau\right) \nonumber\\
&&=\left(\tau^{0 i} \beta^{j} \tau+\tau^{0 j} \beta^{i} \tau\right) \nonumber \\ &&+\tau^{00}\left(x^{\prime i} \beta^{j}+x^{\prime j} \beta^{i}-2 \beta^{i} \beta^{j} \tau\right) ~.
\end{eqnarray}
Then, the second order derivatives are
\begin{eqnarray}
&&\partial_{00} \tau^{00} \left(\left(x^{\prime i}-\beta^{i} \tau\right) \beta^{j} \tau+\left(x^{\prime j}-\beta^{j} \tau\right) \beta^{i} \tau\right)
\nonumber\\
&&+ \partial_{00} \left( \tau^{00} \beta^{i} \beta^{j} \tau^{2}\right) \nonumber\\
&&=\partial_{0}\left(\tau^{0 i} \beta^{j} \tau+\tau^{0 j} \beta^{i} \tau \right) \nonumber\\
&&+\partial_{0}\left(\tau^{00}\left(x^{\prime i} \beta^{j}+x^{\prime j} \beta^{i}-2 \beta^{i} \beta^{j} \tau\right)\right)
\nonumber \\
&&=2 \beta^{j} \tau^{0 i}+2 \beta^{i} \tau^{0 j}-2 \beta^{i} \beta^{j} \tau^{00} ~. \label{eq:B3}
\end{eqnarray}

We notice that the last two terms of Eq.~(\ref{eq:B1}) actually give a constant part after space integral, because their equivalent form (i.e. Eq.~(\ref{eq:B3})) are simply proportional to the total mass energy ($\int \tau^{0 0}$) or momentum ($\int \tau^{0 i}$) of the source. As a result, Eq.~(\ref{eq:9}) is derived and we also calculate $Q^{0}$ in Eq.~(\ref{eq:8}) here for completeness,
\begin{eqnarray}
Q^{0}(\tau) &&=4 \times \frac{1}{2}\left(\frac{\mathrm{d}}{\mathrm{d} \tau}\right)^{2} \int \tau^{00}\left(x^{\prime i}-\beta^{i} \tau\right)\left(x^{\prime j}-\beta^{j} \tau\right) \mathrm{d}^{3} x^{\prime} \nonumber\\
&&=2\left(\frac{\mathrm{d}}{\mathrm{d} \tau}\right)^{2}\left[\sum_{m} \gamma_{m}(\tau) M_{m} \delta_{m}^{i}(\tau) \delta_{m}^{j}(\tau)\right] ~.
\end{eqnarray}


\bibliography{apscit}

\providecommand{\noopsort}[1]{}\providecommand{\singleletter}[1]{#1}%
\begin{thebibliography}{39}%
\makeatletter
\providecommand \@ifxundefined [1]{%
 \@ifx{#1\undefined}
}%
\providecommand \@ifnum [1]{%
 \ifnum #1\expandafter \@firstoftwo
 \else \expandafter \@secondoftwo
 \fi
}%
\providecommand \@ifx [1]{%
 \ifx #1\expandafter \@firstoftwo
 \else \expandafter \@secondoftwo
 \fi
}%
\providecommand \natexlab [1]{#1}%
\providecommand \enquote  [1]{``#1''}%
\providecommand \bibnamefont  [1]{#1}%
\providecommand \bibfnamefont [1]{#1}%
\providecommand \citenamefont [1]{#1}%
\providecommand \href@noop [0]{\@secondoftwo}%
\providecommand \href [0]{\begingroup \@sanitize@url \@href}%
\providecommand \@href[1]{\@@startlink{#1}\@@href}%
\providecommand \@@href[1]{\endgroup#1\@@endlink}%
\providecommand \@sanitize@url [0]{\catcode `\\12\catcode `\$12\catcode
  `\&12\catcode `\#12\catcode `\^12\catcode `\_12\catcode `\%12\relax}%
\providecommand \@@startlink[1]{}%
\providecommand \@@endlink[0]{}%
\providecommand \url  [0]{\begingroup\@sanitize@url \@url }%
\providecommand \@url [1]{\endgroup\@href {#1}{\urlprefix }}%
\providecommand \urlprefix  [0]{URL }%
\providecommand \Eprint [0]{\href }%
\providecommand \doibase [0]{https://doi.org/}%
\providecommand \selectlanguage [0]{\@gobble}%
\providecommand \bibinfo  [0]{\@secondoftwo}%
\providecommand \bibfield  [0]{\@secondoftwo}%
\providecommand \translation [1]{[#1]}%
\providecommand \BibitemOpen [0]{}%
\providecommand \bibitemStop [0]{}%
\providecommand \bibitemNoStop [0]{.\EOS\space}%
\providecommand \EOS [0]{\spacefactor3000\relax}%
\providecommand \BibitemShut  [1]{\csname bibitem#1\endcsname}%
\let\auto@bib@innerbib\@empty
\bibitem [{\citenamefont {{McKernan}}\ \emph {et~al.}(2012)\citenamefont
  {{McKernan}}, \citenamefont {{Ford}}, \citenamefont {{Lyra}},\ and\
  \citenamefont {{Perets}}}]{2012MNRAS.425..460M}%
  \BibitemOpen
  \bibfield  {author} {\bibinfo {author} {\bibfnamefont {B.}~\bibnamefont
  {{McKernan}}}, \bibinfo {author} {\bibfnamefont {K.~E.~S.}\ \bibnamefont
  {{Ford}}}, \bibinfo {author} {\bibfnamefont {W.}~\bibnamefont {{Lyra}}},\
  and\ \bibinfo {author} {\bibfnamefont {H.~B.}\ \bibnamefont {{Perets}}},\
  }\bibfield  {title} {\bibinfo {title} {{Intermediate mass black holes in AGN
  discs - I. Production and growth}},\ }\href
  {https://doi.org/10.1111/j.1365-2966.2012.21486.x} {\bibfield  {journal}
  {\bibinfo  {journal} {Monthly Notices of the Royal Astronomical Society}\
  }\textbf {\bibinfo {volume} {425}},\ \bibinfo {pages} {460} (\bibinfo {year}
  {2012})},\ \Eprint {https://arxiv.org/abs/1206.2309} {arXiv:1206.2309
  [astro-ph.GA]} \BibitemShut {NoStop}%
\bibitem [{\citenamefont {{Secunda}}\ \emph {et~al.}(2019)\citenamefont
  {{Secunda}}, \citenamefont {{Bellovary}}, \citenamefont {{Mac Low}},
  \citenamefont {{Ford}}, \citenamefont {{McKernan}}, \citenamefont {{Leigh}},
  \citenamefont {{Lyra}},\ and\ \citenamefont
  {{S{\'a}ndor}}}]{2019ApJ...878...85S}%
  \BibitemOpen
  \bibfield  {author} {\bibinfo {author} {\bibfnamefont {A.}~\bibnamefont
  {{Secunda}}}, \bibinfo {author} {\bibfnamefont {J.}~\bibnamefont
  {{Bellovary}}}, \bibinfo {author} {\bibfnamefont {M.-M.}\ \bibnamefont {{Mac
  Low}}}, \bibinfo {author} {\bibfnamefont {K.~E.~S.}\ \bibnamefont {{Ford}}},
  \bibinfo {author} {\bibfnamefont {B.}~\bibnamefont {{McKernan}}}, \bibinfo
  {author} {\bibfnamefont {N.~W.~C.}\ \bibnamefont {{Leigh}}}, \bibinfo
  {author} {\bibfnamefont {W.}~\bibnamefont {{Lyra}}},\ and\ \bibinfo {author}
  {\bibfnamefont {Z.}~\bibnamefont {{S{\'a}ndor}}},\ }\bibfield  {title}
  {\bibinfo {title} {{Orbital Migration of Interacting Stellar Mass Black Holes
  in Disks around Supermassive Black Holes}},\ }\href
  {https://doi.org/10.3847/1538-4357/ab20ca} {\bibfield  {journal} {\bibinfo
  {journal} {\apj}\ }\textbf {\bibinfo {volume} {878}},\ \bibinfo {eid} {85}
  (\bibinfo {year} {2019})},\ \Eprint {https://arxiv.org/abs/1807.02859}
  {arXiv:1807.02859 [astro-ph.HE]} \BibitemShut {NoStop}%
\bibitem [{\citenamefont {{Tagawa}}\ \emph {et~al.}(2020)\citenamefont
  {{Tagawa}}, \citenamefont {{Haiman}},\ and\ \citenamefont
  {{Kocsis}}}]{2020ApJ...898...25T}%
  \BibitemOpen
  \bibfield  {author} {\bibinfo {author} {\bibfnamefont {H.}~\bibnamefont
  {{Tagawa}}}, \bibinfo {author} {\bibfnamefont {Z.}~\bibnamefont {{Haiman}}},\
  and\ \bibinfo {author} {\bibfnamefont {B.}~\bibnamefont {{Kocsis}}},\
  }\bibfield  {title} {\bibinfo {title} {{Formation and Evolution of
  Compact-object Binaries in AGN Disks}},\ }\href
  {https://doi.org/10.3847/1538-4357/ab9b8c} {\bibfield  {journal} {\bibinfo
  {journal} {\apj}\ }\textbf {\bibinfo {volume} {898}},\ \bibinfo {eid} {25}
  (\bibinfo {year} {2020})},\ \Eprint {https://arxiv.org/abs/1912.08218}
  {arXiv:1912.08218 [astro-ph.GA]} \BibitemShut {NoStop}%
\bibitem [{\citenamefont {{McKernan}}\ \emph {et~al.}(2018)\citenamefont
  {{McKernan}}, \citenamefont {{Ford}}, \citenamefont {{Bellovary}},
  \citenamefont {{Leigh}}, \citenamefont {{Haiman}}, \citenamefont {{Kocsis}},
  \citenamefont {{Lyra}}, \citenamefont {{Mac Low}}, \citenamefont {{Metzger}},
  \citenamefont {{O'Dowd}}, \citenamefont {{Endlich}},\ and\ \citenamefont
  {{Rosen}}}]{2018ApJ...866...66M}%
  \BibitemOpen
  \bibfield  {author} {\bibinfo {author} {\bibfnamefont {B.}~\bibnamefont
  {{McKernan}}}, \bibinfo {author} {\bibfnamefont {K.~E.~S.}\ \bibnamefont
  {{Ford}}}, \bibinfo {author} {\bibfnamefont {J.}~\bibnamefont {{Bellovary}}},
  \bibinfo {author} {\bibfnamefont {N.~W.~C.}\ \bibnamefont {{Leigh}}},
  \bibinfo {author} {\bibfnamefont {Z.}~\bibnamefont {{Haiman}}}, \bibinfo
  {author} {\bibfnamefont {B.}~\bibnamefont {{Kocsis}}}, \bibinfo {author}
  {\bibfnamefont {W.}~\bibnamefont {{Lyra}}}, \bibinfo {author} {\bibfnamefont
  {M.~M.}\ \bibnamefont {{Mac Low}}}, \bibinfo {author} {\bibfnamefont
  {B.}~\bibnamefont {{Metzger}}}, \bibinfo {author} {\bibfnamefont
  {M.}~\bibnamefont {{O'Dowd}}}, \bibinfo {author} {\bibfnamefont
  {S.}~\bibnamefont {{Endlich}}},\ and\ \bibinfo {author} {\bibfnamefont
  {D.~J.}\ \bibnamefont {{Rosen}}},\ }\bibfield  {title} {\bibinfo {title}
  {{Constraining Stellar-mass Black Hole Mergers in AGN Disks Detectable with
  LIGO}},\ }\href {https://doi.org/10.3847/1538-4357/aadae5} {\bibfield
  {journal} {\bibinfo  {journal} {\apj}\ }\textbf {\bibinfo {volume} {866}},\
  \bibinfo {eid} {66} (\bibinfo {year} {2018})},\ \Eprint
  {https://arxiv.org/abs/1702.07818} {arXiv:1702.07818 [astro-ph.HE]}
  \BibitemShut {NoStop}%
\bibitem [{\citenamefont {{Antoni}}\ \emph {et~al.}(2019)\citenamefont
  {{Antoni}}, \citenamefont {{MacLeod}},\ and\ \citenamefont
  {{Ramirez-Ruiz}}}]{2019ApJ...884...22A}%
  \BibitemOpen
  \bibfield  {author} {\bibinfo {author} {\bibfnamefont {A.}~\bibnamefont
  {{Antoni}}}, \bibinfo {author} {\bibfnamefont {M.}~\bibnamefont
  {{MacLeod}}},\ and\ \bibinfo {author} {\bibfnamefont {E.}~\bibnamefont
  {{Ramirez-Ruiz}}},\ }\bibfield  {title} {\bibinfo {title} {{The Evolution of
  Binaries in a Gaseous Medium: Three-dimensional Simulations of Binary
  Bondi-Hoyle-Lyttleton Accretion}},\ }\href
  {https://doi.org/10.3847/1538-4357/ab3466} {\bibfield  {journal} {\bibinfo
  {journal} {\apj}\ }\textbf {\bibinfo {volume} {884}},\ \bibinfo {eid} {22}
  (\bibinfo {year} {2019})},\ \Eprint {https://arxiv.org/abs/1901.07572}
  {arXiv:1901.07572 [astro-ph.HE]} \BibitemShut {NoStop}%
\bibitem [{\citenamefont {{Secunda}}\ \emph {et~al.}(2020)\citenamefont
  {{Secunda}}, \citenamefont {{Bellovary}}, \citenamefont {{Mac Low}},
  \citenamefont {{Ford}}, \citenamefont {{McKernan}}, \citenamefont {{Leigh}},
  \citenamefont {{Lyra}}, \citenamefont {{S{\'a}ndor}},\ and\ \citenamefont
  {{Adorno}}}]{2020ApJ...903..133S}%
  \BibitemOpen
  \bibfield  {author} {\bibinfo {author} {\bibfnamefont {A.}~\bibnamefont
  {{Secunda}}}, \bibinfo {author} {\bibfnamefont {J.}~\bibnamefont
  {{Bellovary}}}, \bibinfo {author} {\bibfnamefont {M.-M.}\ \bibnamefont {{Mac
  Low}}}, \bibinfo {author} {\bibfnamefont {K.~E.~S.}\ \bibnamefont {{Ford}}},
  \bibinfo {author} {\bibfnamefont {B.}~\bibnamefont {{McKernan}}}, \bibinfo
  {author} {\bibfnamefont {N.~W.~C.}\ \bibnamefont {{Leigh}}}, \bibinfo
  {author} {\bibfnamefont {W.}~\bibnamefont {{Lyra}}}, \bibinfo {author}
  {\bibfnamefont {Z.}~\bibnamefont {{S{\'a}ndor}}},\ and\ \bibinfo {author}
  {\bibfnamefont {J.~I.}\ \bibnamefont {{Adorno}}},\ }\bibfield  {title}
  {\bibinfo {title} {{Orbital Migration of Interacting Stellar Mass Black Holes
  in Disks around Supermassive Black Holes. II. Spins and Incoming Objects}},\
  }\href {https://doi.org/10.3847/1538-4357/abbc1d} {\bibfield  {journal}
  {\bibinfo  {journal} {\apj}\ }\textbf {\bibinfo {volume} {903}},\ \bibinfo
  {eid} {133} (\bibinfo {year} {2020})},\ \Eprint
  {https://arxiv.org/abs/2004.11936} {arXiv:2004.11936 [astro-ph.HE]}
  \BibitemShut {NoStop}%
\bibitem [{\citenamefont {{Gr{\"o}bner}}\ \emph {et~al.}(2020)\citenamefont
  {{Gr{\"o}bner}}, \citenamefont {{Ishibashi}}, \citenamefont {{Tiwari}},
  \citenamefont {{Haney}},\ and\ \citenamefont
  {{Jetzer}}}]{2020A&A...638A.119G}%
  \BibitemOpen
  \bibfield  {author} {\bibinfo {author} {\bibfnamefont {M.}~\bibnamefont
  {{Gr{\"o}bner}}}, \bibinfo {author} {\bibfnamefont {W.}~\bibnamefont
  {{Ishibashi}}}, \bibinfo {author} {\bibfnamefont {S.}~\bibnamefont
  {{Tiwari}}}, \bibinfo {author} {\bibfnamefont {M.}~\bibnamefont {{Haney}}},\
  and\ \bibinfo {author} {\bibfnamefont {P.}~\bibnamefont {{Jetzer}}},\
  }\bibfield  {title} {\bibinfo {title} {{Binary black hole mergers in AGN
  accretion discs: gravitational wave rate density estimates}},\ }\href
  {https://doi.org/10.1051/0004-6361/202037681} {\bibfield  {journal} {\bibinfo
   {journal} {Astronomy \& Astrophysics}\ }\textbf {\bibinfo {volume} {638}},\
  \bibinfo {eid} {A119} (\bibinfo {year} {2020})},\ \Eprint
  {https://arxiv.org/abs/2005.03571} {arXiv:2005.03571 [astro-ph.GA]}
  \BibitemShut {NoStop}%
\bibitem [{\citenamefont {{Chen}}\ \emph {et~al.}(2019)\citenamefont {{Chen}},
  \citenamefont {{Li}},\ and\ \citenamefont {{Cao}}}]{2019MNRAS.485L.141C}%
  \BibitemOpen
  \bibfield  {author} {\bibinfo {author} {\bibfnamefont {X.}~\bibnamefont
  {{Chen}}}, \bibinfo {author} {\bibfnamefont {S.}~\bibnamefont {{Li}}},\ and\
  \bibinfo {author} {\bibfnamefont {Z.}~\bibnamefont {{Cao}}},\ }\bibfield
  {title} {\bibinfo {title} {{Mass-redshift degeneracy for the
  gravitational-wave sources in the vicinity of supermassive black holes}},\
  }\href {https://doi.org/10.1093/mnrasl/slz046} {\bibfield  {journal}
  {\bibinfo  {journal} {Monthly Notices of the Royal Astronomical Society:
  Letters}\ }\textbf {\bibinfo {volume} {485}},\ \bibinfo {pages} {L141}
  (\bibinfo {year} {2019})},\ \Eprint {https://arxiv.org/abs/1703.10543}
  {arXiv:1703.10543 [astro-ph.HE]} \BibitemShut {NoStop}%
\bibitem [{\citenamefont {{Peng}}\ and\ \citenamefont
  {{Chen}}(2021)}]{2021MNRAS.505.1324P}%
  \BibitemOpen
  \bibfield  {author} {\bibinfo {author} {\bibfnamefont {P.}~\bibnamefont
  {{Peng}}}\ and\ \bibinfo {author} {\bibfnamefont {X.}~\bibnamefont
  {{Chen}}},\ }\bibfield  {title} {\bibinfo {title} {{The last migration trap
  of compact objects in AGN accretion disc}},\ }\href
  {https://doi.org/10.1093/mnras/stab1419} {\bibfield  {journal} {\bibinfo
  {journal} {Monthly Notices of the Royal Astronomical Society}\ }\textbf
  {\bibinfo {volume} {505}},\ \bibinfo {pages} {1324} (\bibinfo {year}
  {2021})},\ \Eprint {https://arxiv.org/abs/2104.07685} {arXiv:2104.07685
  [astro-ph.HE]} \BibitemShut {NoStop}%
\bibitem [{\citenamefont {{Bonvin}}\ \emph {et~al.}(2017)\citenamefont
  {{Bonvin}}, \citenamefont {{Caprini}}, \citenamefont {{Sturani}},\ and\
  \citenamefont {{Tamanini}}}]{2017PhRvD..95d4029B}%
  \BibitemOpen
  \bibfield  {author} {\bibinfo {author} {\bibfnamefont {C.}~\bibnamefont
  {{Bonvin}}}, \bibinfo {author} {\bibfnamefont {C.}~\bibnamefont {{Caprini}}},
  \bibinfo {author} {\bibfnamefont {R.}~\bibnamefont {{Sturani}}},\ and\
  \bibinfo {author} {\bibfnamefont {N.}~\bibnamefont {{Tamanini}}},\ }\bibfield
   {title} {\bibinfo {title} {{Effect of matter structure on the gravitational
  waveform}},\ }\href {https://doi.org/10.1103/PhysRevD.95.044029} {\bibfield
  {journal} {\bibinfo  {journal} {\prd}\ }\textbf {\bibinfo {volume} {95}},\
  \bibinfo {eid} {044029} (\bibinfo {year} {2017})},\ \Eprint
  {https://arxiv.org/abs/1609.08093} {arXiv:1609.08093 [astro-ph.CO]}
  \BibitemShut {NoStop}%
\bibitem [{\citenamefont {{Chen}}\ \emph {et~al.}(2020)\citenamefont {{Chen}},
  \citenamefont {{Xuan}},\ and\ \citenamefont {{Peng}}}]{2020ApJ...896..171C}%
  \BibitemOpen
  \bibfield  {author} {\bibinfo {author} {\bibfnamefont {X.}~\bibnamefont
  {{Chen}}}, \bibinfo {author} {\bibfnamefont {Z.-Y.}\ \bibnamefont {{Xuan}}},\
  and\ \bibinfo {author} {\bibfnamefont {P.}~\bibnamefont {{Peng}}},\
  }\bibfield  {title} {\bibinfo {title} {{Fake Massive Black Holes in the
  Milli-Hertz Gravitational-wave Band}},\ }\href
  {https://doi.org/10.3847/1538-4357/ab919f} {\bibfield  {journal} {\bibinfo
  {journal} {\apj}\ }\textbf {\bibinfo {volume} {896}},\ \bibinfo {eid} {171}
  (\bibinfo {year} {2020})},\ \Eprint {https://arxiv.org/abs/2003.08639}
  {arXiv:2003.08639 [astro-ph.HE]} \BibitemShut {NoStop}%
\bibitem [{\citenamefont {{Meiron}}\ \emph {et~al.}(2017)\citenamefont
  {{Meiron}}, \citenamefont {{Kocsis}},\ and\ \citenamefont
  {{Loeb}}}]{2017ApJ...834..200M}%
  \BibitemOpen
  \bibfield  {author} {\bibinfo {author} {\bibfnamefont {Y.}~\bibnamefont
  {{Meiron}}}, \bibinfo {author} {\bibfnamefont {B.}~\bibnamefont {{Kocsis}}},\
  and\ \bibinfo {author} {\bibfnamefont {A.}~\bibnamefont {{Loeb}}},\
  }\bibfield  {title} {\bibinfo {title} {{Detecting Triple Systems with
  Gravitational Wave Observations}},\ }\href
  {https://doi.org/10.3847/1538-4357/834/2/200} {\bibfield  {journal} {\bibinfo
   {journal} {\apj}\ }\textbf {\bibinfo {volume} {834}},\ \bibinfo {eid} {200}
  (\bibinfo {year} {2017})},\ \Eprint {https://arxiv.org/abs/1604.02148}
  {arXiv:1604.02148 [astro-ph.HE]} \BibitemShut {NoStop}%
\bibitem [{\citenamefont {{Inayoshi}}\ \emph {et~al.}(2017)\citenamefont
  {{Inayoshi}}, \citenamefont {{Tamanini}}, \citenamefont {{Caprini}},\ and\
  \citenamefont {{Haiman}}}]{2017PhRvD..96f3014I}%
  \BibitemOpen
  \bibfield  {author} {\bibinfo {author} {\bibfnamefont {K.}~\bibnamefont
  {{Inayoshi}}}, \bibinfo {author} {\bibfnamefont {N.}~\bibnamefont
  {{Tamanini}}}, \bibinfo {author} {\bibfnamefont {C.}~\bibnamefont
  {{Caprini}}},\ and\ \bibinfo {author} {\bibfnamefont {Z.}~\bibnamefont
  {{Haiman}}},\ }\bibfield  {title} {\bibinfo {title} {{Probing stellar binary
  black hole formation in galactic nuclei via the imprint of their center of
  mass acceleration on their gravitational wave signal}},\ }\href
  {https://doi.org/10.1103/PhysRevD.96.063014} {\bibfield  {journal} {\bibinfo
  {journal} {\prd}\ }\textbf {\bibinfo {volume} {96}},\ \bibinfo {eid} {063014}
  (\bibinfo {year} {2017})},\ \Eprint {https://arxiv.org/abs/1702.06529}
  {arXiv:1702.06529 [astro-ph.HE]} \BibitemShut {NoStop}%
\bibitem [{\citenamefont {{Wong}}\ \emph {et~al.}(2019)\citenamefont {{Wong}},
  \citenamefont {{Baibhav}},\ and\ \citenamefont
  {{Berti}}}]{2019MNRAS.488.5665W}%
  \BibitemOpen
  \bibfield  {author} {\bibinfo {author} {\bibfnamefont {K.~W.~K.}\
  \bibnamefont {{Wong}}}, \bibinfo {author} {\bibfnamefont {V.}~\bibnamefont
  {{Baibhav}}},\ and\ \bibinfo {author} {\bibfnamefont {E.}~\bibnamefont
  {{Berti}}},\ }\bibfield  {title} {\bibinfo {title} {{Binary radial velocity
  measurements with space-based gravitational-wave detectors}},\ }\href
  {https://doi.org/10.1093/mnras/stz2077} {\bibfield  {journal} {\bibinfo
  {journal} {Monthly Notices of the Royal Astronomical Society}\ }\textbf
  {\bibinfo {volume} {488}},\ \bibinfo {pages} {5665} (\bibinfo {year}
  {2019})},\ \Eprint {https://arxiv.org/abs/1902.01402} {arXiv:1902.01402
  [astro-ph.HE]} \BibitemShut {NoStop}%
\bibitem [{\citenamefont {{Woodford}}\ \emph {et~al.}(2019)\citenamefont
  {{Woodford}}, \citenamefont {{Boyle}},\ and\ \citenamefont
  {{Pfeiffer}}}]{2019PhRvD.100l4010W}%
  \BibitemOpen
  \bibfield  {author} {\bibinfo {author} {\bibfnamefont {C.~J.}\ \bibnamefont
  {{Woodford}}}, \bibinfo {author} {\bibfnamefont {M.}~\bibnamefont
  {{Boyle}}},\ and\ \bibinfo {author} {\bibfnamefont {H.~P.}\ \bibnamefont
  {{Pfeiffer}}},\ }\bibfield  {title} {\bibinfo {title} {{Compact binary
  waveform center-of-mass corrections}},\ }\href
  {https://doi.org/10.1103/PhysRevD.100.124010} {\bibfield  {journal} {\bibinfo
   {journal} {\prd}\ }\textbf {\bibinfo {volume} {100}},\ \bibinfo {eid}
  {124010} (\bibinfo {year} {2019})},\ \Eprint
  {https://arxiv.org/abs/1904.04842} {arXiv:1904.04842 [gr-qc]} \BibitemShut
  {NoStop}%
\bibitem [{\citenamefont {{Randall}}\ and\ \citenamefont
  {{Xianyu}}(2019)}]{2019ApJ...878...75R}%
  \BibitemOpen
  \bibfield  {author} {\bibinfo {author} {\bibfnamefont {L.}~\bibnamefont
  {{Randall}}}\ and\ \bibinfo {author} {\bibfnamefont {Z.-Z.}\ \bibnamefont
  {{Xianyu}}},\ }\bibfield  {title} {\bibinfo {title} {{A Direct Probe of Mass
  Density near Inspiraling Binary Black Holes}},\ }\href
  {https://doi.org/10.3847/1538-4357/ab20c6} {\bibfield  {journal} {\bibinfo
  {journal} {\apj}\ }\textbf {\bibinfo {volume} {878}},\ \bibinfo {eid} {75}
  (\bibinfo {year} {2019})},\ \Eprint {https://arxiv.org/abs/1805.05335}
  {arXiv:1805.05335 [gr-qc]} \BibitemShut {NoStop}%
\bibitem [{\citenamefont {{Torres-Orjuela}}\ \emph {et~al.}(2020)\citenamefont
  {{Torres-Orjuela}}, \citenamefont {{Chen}},\ and\ \citenamefont
  {{Amaro-Seoane}}}]{2020PhRvD.101h3028T}%
  \BibitemOpen
  \bibfield  {author} {\bibinfo {author} {\bibfnamefont {A.}~\bibnamefont
  {{Torres-Orjuela}}}, \bibinfo {author} {\bibfnamefont {X.}~\bibnamefont
  {{Chen}}},\ and\ \bibinfo {author} {\bibfnamefont {P.}~\bibnamefont
  {{Amaro-Seoane}}},\ }\bibfield  {title} {\bibinfo {title} {{Phase shift of
  gravitational waves induced by aberration}},\ }\href
  {https://doi.org/10.1103/PhysRevD.101.083028} {\bibfield  {journal} {\bibinfo
   {journal} {\prd}\ }\textbf {\bibinfo {volume} {101}},\ \bibinfo {eid}
  {083028} (\bibinfo {year} {2020})},\ \Eprint
  {https://arxiv.org/abs/2001.00721} {arXiv:2001.00721 [astro-ph.HE]}
  \BibitemShut {NoStop}%
\bibitem [{\citenamefont {{Tamanini}}\ \emph {et~al.}(2020)\citenamefont
  {{Tamanini}}, \citenamefont {{Klein}}, \citenamefont {{Bonvin}},
  \citenamefont {{Barausse}},\ and\ \citenamefont
  {{Caprini}}}]{2020PhRvD.101f3002T}%
  \BibitemOpen
  \bibfield  {author} {\bibinfo {author} {\bibfnamefont {N.}~\bibnamefont
  {{Tamanini}}}, \bibinfo {author} {\bibfnamefont {A.}~\bibnamefont {{Klein}}},
  \bibinfo {author} {\bibfnamefont {C.}~\bibnamefont {{Bonvin}}}, \bibinfo
  {author} {\bibfnamefont {E.}~\bibnamefont {{Barausse}}},\ and\ \bibinfo
  {author} {\bibfnamefont {C.}~\bibnamefont {{Caprini}}},\ }\bibfield  {title}
  {\bibinfo {title} {{Peculiar acceleration of stellar-origin black hole
  binaries: Measurement and biases with LISA}},\ }\href
  {https://doi.org/10.1103/PhysRevD.101.063002} {\bibfield  {journal} {\bibinfo
   {journal} {\prd}\ }\textbf {\bibinfo {volume} {101}},\ \bibinfo {eid}
  {063002} (\bibinfo {year} {2020})},\ \Eprint
  {https://arxiv.org/abs/1907.02018} {arXiv:1907.02018 [astro-ph.IM]}
  \BibitemShut {NoStop}%
\bibitem [{\citenamefont {{Yu}}\ and\ \citenamefont
  {{Chen}}(2021)}]{2021PhRvL.126b1101Y}%
  \BibitemOpen
  \bibfield  {author} {\bibinfo {author} {\bibfnamefont {H.}~\bibnamefont
  {{Yu}}}\ and\ \bibinfo {author} {\bibfnamefont {Y.}~\bibnamefont {{Chen}}},\
  }\bibfield  {title} {\bibinfo {title} {{Direct Determination of Supermassive
  Black Hole Properties with Gravitational-Wave Radiation from Surrounding
  Stellar-Mass Black Hole Binaries}},\ }\href
  {https://doi.org/10.1103/PhysRevLett.126.021101} {\bibfield  {journal}
  {\bibinfo  {journal} {\prl}\ }\textbf {\bibinfo {volume} {126}},\ \bibinfo
  {eid} {021101} (\bibinfo {year} {2021})},\ \Eprint
  {https://arxiv.org/abs/2009.02579} {arXiv:2009.02579 [gr-qc]} \BibitemShut
  {NoStop}%
\bibitem [{\citenamefont {{Chamberlain}}\ \emph {et~al.}(2019)\citenamefont
  {{Chamberlain}}, \citenamefont {{Moore}}, \citenamefont {{Gerosa}},\ and\
  \citenamefont {{Yunes}}}]{2019PhRvD..99b4025C}%
  \BibitemOpen
  \bibfield  {author} {\bibinfo {author} {\bibfnamefont {K.}~\bibnamefont
  {{Chamberlain}}}, \bibinfo {author} {\bibfnamefont {C.~J.}\ \bibnamefont
  {{Moore}}}, \bibinfo {author} {\bibfnamefont {D.}~\bibnamefont {{Gerosa}}},\
  and\ \bibinfo {author} {\bibfnamefont {N.}~\bibnamefont {{Yunes}}},\
  }\bibfield  {title} {\bibinfo {title} {{Frequency-domain waveform
  approximants capturing Doppler shifts}},\ }\href
  {https://doi.org/10.1103/PhysRevD.99.024025} {\bibfield  {journal} {\bibinfo
  {journal} {\prd}\ }\textbf {\bibinfo {volume} {99}},\ \bibinfo {eid} {024025}
  (\bibinfo {year} {2019})},\ \Eprint {https://arxiv.org/abs/1809.04799}
  {arXiv:1809.04799 [gr-qc]} \BibitemShut {NoStop}%
\bibitem [{\citenamefont {{Xuan}}\ \emph {et~al.}(2021)\citenamefont {{Xuan}},
  \citenamefont {{Peng}},\ and\ \citenamefont {{Chen}}}]{2021MNRAS.502.4199X}%
  \BibitemOpen
  \bibfield  {author} {\bibinfo {author} {\bibfnamefont {Z.}~\bibnamefont
  {{Xuan}}}, \bibinfo {author} {\bibfnamefont {P.}~\bibnamefont {{Peng}}},\
  and\ \bibinfo {author} {\bibfnamefont {X.}~\bibnamefont {{Chen}}},\
  }\bibfield  {title} {\bibinfo {title} {{Degeneracy between mass and peculiar
  acceleration for the double white dwarfs in the LISA band}},\ }\href
  {https://doi.org/10.1093/mnras/stab331} {\bibfield  {journal} {\bibinfo
  {journal} {Monthly Notices of the Royal Astronomical Society}\ }\textbf
  {\bibinfo {volume} {502}},\ \bibinfo {pages} {4199} (\bibinfo {year}
  {2021})},\ \Eprint {https://arxiv.org/abs/2012.00049} {arXiv:2012.00049
  [astro-ph.HE]} \BibitemShut {NoStop}%
\bibitem [{\citenamefont {{Peters}}\ and\ \citenamefont
  {{Mathews}}(1963)}]{1963PhRv..131..435P}%
  \BibitemOpen
  \bibfield  {author} {\bibinfo {author} {\bibfnamefont {P.~C.}\ \bibnamefont
  {{Peters}}}\ and\ \bibinfo {author} {\bibfnamefont {J.}~\bibnamefont
  {{Mathews}}},\ }\bibfield  {title} {\bibinfo {title} {{Gravitational
  Radiation from Point Masses in a Keplerian Orbit}},\ }\href
  {https://doi.org/10.1103/PhysRev.131.435} {\bibfield  {journal} {\bibinfo
  {journal} {Physical Review}\ }\textbf {\bibinfo {volume} {131}},\ \bibinfo
  {pages} {435} (\bibinfo {year} {1963})}\BibitemShut {NoStop}%
\bibitem [{\citenamefont {{Peters}}(1964)}]{1964PhRv..136.1224P}%
  \BibitemOpen
  \bibfield  {author} {\bibinfo {author} {\bibfnamefont {P.~C.}\ \bibnamefont
  {{Peters}}},\ }\bibfield  {title} {\bibinfo {title} {{Gravitational Radiation
  and the Motion of Two Point Masses}},\ }\href
  {https://doi.org/10.1103/PhysRev.136.B1224} {\bibfield  {journal} {\bibinfo
  {journal} {Physical Review}\ }\textbf {\bibinfo {volume} {136}},\ \bibinfo
  {pages} {1224} (\bibinfo {year} {1964})}\BibitemShut {NoStop}%
\bibitem [{\citenamefont {{Torres-Orjuela}}\ \emph {et~al.}(2021)\citenamefont
  {{Torres-Orjuela}}, \citenamefont {{Amaro Seoane}}, \citenamefont {{Xuan}},
  \citenamefont {{Chua}}, \citenamefont {{Rosell}},\ and\ \citenamefont
  {{Chen}}}]{2021PhRvL.127d1102T}%
  \BibitemOpen
  \bibfield  {author} {\bibinfo {author} {\bibfnamefont {A.}~\bibnamefont
  {{Torres-Orjuela}}}, \bibinfo {author} {\bibfnamefont {P.}~\bibnamefont
  {{Amaro Seoane}}}, \bibinfo {author} {\bibfnamefont {Z.}~\bibnamefont
  {{Xuan}}}, \bibinfo {author} {\bibfnamefont {A.~J.~K.}\ \bibnamefont
  {{Chua}}}, \bibinfo {author} {\bibfnamefont {M.~J.~B.}\ \bibnamefont
  {{Rosell}}},\ and\ \bibinfo {author} {\bibfnamefont {X.}~\bibnamefont
  {{Chen}}},\ }\bibfield  {title} {\bibinfo {title} {{Exciting Modes due to the
  Aberration of Gravitational Waves: Measurability for Extreme-Mass-Ratio
  Inspirals}},\ }\href {https://doi.org/10.1103/PhysRevLett.127.041102}
  {\bibfield  {journal} {\bibinfo  {journal} {\prl}\ }\textbf {\bibinfo
  {volume} {127}},\ \bibinfo {eid} {041102} (\bibinfo {year} {2021})},\ \Eprint
  {https://arxiv.org/abs/2010.15842} {arXiv:2010.15842 [gr-qc]} \BibitemShut
  {NoStop}%
\bibitem [{\citenamefont {{Torres-Orjuela}}\ \emph {et~al.}(2019)\citenamefont
  {{Torres-Orjuela}}, \citenamefont {{Chen}}, \citenamefont {{Cao}},
  \citenamefont {{Amaro-Seoane}},\ and\ \citenamefont
  {{Peng}}}]{2019PhRvD.100f3012T}%
  \BibitemOpen
  \bibfield  {author} {\bibinfo {author} {\bibfnamefont {A.}~\bibnamefont
  {{Torres-Orjuela}}}, \bibinfo {author} {\bibfnamefont {X.}~\bibnamefont
  {{Chen}}}, \bibinfo {author} {\bibfnamefont {Z.}~\bibnamefont {{Cao}}},
  \bibinfo {author} {\bibfnamefont {P.}~\bibnamefont {{Amaro-Seoane}}},\ and\
  \bibinfo {author} {\bibfnamefont {P.}~\bibnamefont {{Peng}}},\ }\bibfield
  {title} {\bibinfo {title} {{Detecting the beaming effect of gravitational
  waves}},\ }\href {https://doi.org/10.1103/PhysRevD.100.063012} {\bibfield
  {journal} {\bibinfo  {journal} {\prd}\ }\textbf {\bibinfo {volume} {100}},\
  \bibinfo {eid} {063012} (\bibinfo {year} {2019})},\ \Eprint
  {https://arxiv.org/abs/1806.09857} {arXiv:1806.09857 [astro-ph.HE]}
  \BibitemShut {NoStop}%
\bibitem [{\citenamefont {{Thorne}}(1983)}]{1983LNP...124....1T}%
  \BibitemOpen
  \bibfield  {author} {\bibinfo {author} {\bibfnamefont {K.~S.}\ \bibnamefont
  {{Thorne}}},\ }\bibfield  {title} {\bibinfo {title} {{The theory of
  gravitational radiation: an introductory review.}},\ }in\ \href@noop {}
  {\emph {\bibinfo {booktitle} {Lecture Notes in Physics, Berlin Springer
  Verlag}}},\ Vol.\ \bibinfo {volume} {124}\ (\bibinfo {year} {1983})\ pp.\
  \bibinfo {pages} {1--57}\BibitemShut {NoStop}%
\bibitem [{\citenamefont {{Hawking}}\ and\ \citenamefont
  {{Israel}}(1987)}]{1987thyg.book.....H}%
  \BibitemOpen
  \bibfield  {author} {\bibinfo {author} {\bibfnamefont {S.~W.}\ \bibnamefont
  {{Hawking}}}\ and\ \bibinfo {author} {\bibfnamefont {W.}~\bibnamefont
  {{Israel}}},\ }\href@noop {} {\emph {\bibinfo {title} {{Three hundred years
  of gravitation}}}}\ (\bibinfo {year} {1987})\BibitemShut {NoStop}%
\bibitem [{\citenamefont {{Husa}}(2009)}]{2009GReGr..41.1667H}%
  \BibitemOpen
  \bibfield  {author} {\bibinfo {author} {\bibfnamefont {S.}~\bibnamefont
  {{Husa}}},\ }\bibfield  {title} {\bibinfo {title} {{Michele Maggiore:
  Gravitational waves. Volume 1: theory and experiments. Oxford University
  Press, 2007, 576p., GBP47.00, ISBN13: 978-0-19-857074-5}},\ }\href
  {https://doi.org/10.1007/s10714-009-0762-5} {\bibfield  {journal} {\bibinfo
  {journal} {General Relativity and Gravitation}\ }\textbf {\bibinfo {volume}
  {41}},\ \bibinfo {pages} {1667} (\bibinfo {year} {2009})}\BibitemShut
  {NoStop}%
\bibitem [{\citenamefont {{Kopeikin}}\ and\ \citenamefont
  {{Sch{\"a}fer}}(1999)}]{1999PhRvD..60l4002K}%
  \BibitemOpen
  \bibfield  {author} {\bibinfo {author} {\bibfnamefont {S.~M.}\ \bibnamefont
  {{Kopeikin}}}\ and\ \bibinfo {author} {\bibfnamefont {G.}~\bibnamefont
  {{Sch{\"a}fer}}},\ }\bibfield  {title} {\bibinfo {title} {{Lorentz covariant
  theory of light propagation in gravitational fields of arbitrary-moving
  bodies}},\ }\href {https://doi.org/10.1103/PhysRevD.60.124002} {\bibfield
  {journal} {\bibinfo  {journal} {\prd}\ }\textbf {\bibinfo {volume} {60}},\
  \bibinfo {eid} {124002} (\bibinfo {year} {1999})},\ \Eprint
  {https://arxiv.org/abs/gr-qc/9902030} {arXiv:gr-qc/9902030 [gr-qc]}
  \BibitemShut {NoStop}%
\bibitem [{\citenamefont {{Kopeikin}}\ and\ \citenamefont
  {{Fomalont}}(2007)}]{2007GReGr..39.1583K}%
  \BibitemOpen
  \bibfield  {author} {\bibinfo {author} {\bibfnamefont {S.~M.}\ \bibnamefont
  {{Kopeikin}}}\ and\ \bibinfo {author} {\bibfnamefont {E.~B.}\ \bibnamefont
  {{Fomalont}}},\ }\bibfield  {title} {\bibinfo {title} {{Gravimagnetism,
  causality, and aberration of gravity in the gravitational light-ray
  deflection experiments}},\ }\href {https://doi.org/10.1007/s10714-007-0483-6}
  {\bibfield  {journal} {\bibinfo  {journal} {General Relativity and
  Gravitation}\ }\textbf {\bibinfo {volume} {39}},\ \bibinfo {pages} {1583}
  (\bibinfo {year} {2007})},\ \Eprint {https://arxiv.org/abs/gr-qc/0510077}
  {arXiv:gr-qc/0510077 [gr-qc]} \BibitemShut {NoStop}%
\bibitem [{\citenamefont {{Zschocke}}\ and\ \citenamefont
  {{Soffel}}(2014)}]{2014CQGra..31q5001Z}%
  \BibitemOpen
  \bibfield  {author} {\bibinfo {author} {\bibfnamefont {S.}~\bibnamefont
  {{Zschocke}}}\ and\ \bibinfo {author} {\bibfnamefont {M.~H.}\ \bibnamefont
  {{Soffel}}},\ }\bibfield  {title} {\bibinfo {title} {{Gravitational field of
  one uniformly moving extended body and N arbitrarily moving pointlike bodies
  in post-Minkowskian approximation}},\ }\href
  {https://doi.org/10.1088/0264-9381/31/17/175001} {\bibfield  {journal}
  {\bibinfo  {journal} {Classical and Quantum Gravity}\ }\textbf {\bibinfo
  {volume} {31}},\ \bibinfo {eid} {175001} (\bibinfo {year} {2014})},\ \Eprint
  {https://arxiv.org/abs/1403.5438} {arXiv:1403.5438 [astro-ph.IM]}
  \BibitemShut {NoStop}%
\bibitem [{\citenamefont {{Press}}(1977)}]{1977PhRvD..15..965P}%
  \BibitemOpen
  \bibfield  {author} {\bibinfo {author} {\bibfnamefont {W.~H.}\ \bibnamefont
  {{Press}}},\ }\bibfield  {title} {\bibinfo {title} {{Gravitational radiation
  from sources which extend into their own wave zone}},\ }\href
  {https://doi.org/10.1103/PhysRevD.15.965} {\bibfield  {journal} {\bibinfo
  {journal} {\prd}\ }\textbf {\bibinfo {volume} {15}},\ \bibinfo {pages} {965}
  (\bibinfo {year} {1977})}\BibitemShut {NoStop}%
\bibitem [{\citenamefont {{Landau}}\ and\ \citenamefont
  {{Lifshitz}}(1975)}]{1975ctf..book.....L}%
  \BibitemOpen
  \bibfield  {author} {\bibinfo {author} {\bibfnamefont {L.~D.}\ \bibnamefont
  {{Landau}}}\ and\ \bibinfo {author} {\bibfnamefont {E.~M.}\ \bibnamefont
  {{Lifshitz}}},\ }\href@noop {} {\emph {\bibinfo {title} {{The classical
  theory of fields}}}}\ (\bibinfo {year} {1975})\BibitemShut {NoStop}%
\bibitem [{\citenamefont {{Poisson}}\ and\ \citenamefont
  {{Will}}(2014)}]{2014grav.book.....P}%
  \BibitemOpen
  \bibfield  {author} {\bibinfo {author} {\bibfnamefont {E.}~\bibnamefont
  {{Poisson}}}\ and\ \bibinfo {author} {\bibfnamefont {C.~M.}\ \bibnamefont
  {{Will}}},\ }\href@noop {} {\emph {\bibinfo {title} {{Gravity}}}}\ (\bibinfo
  {year} {2014})\BibitemShut {NoStop}%
\bibitem [{Note1()}]{Note1}%
  \BibitemOpen
  \bibinfo {note} {The code used for this calculation can be found at
  https://github.com/StrelitziaHY/GW-transformation}\BibitemShut {NoStop}%
\bibitem [{\citenamefont {{Dymnikova}}\ \emph {et~al.}(1982)\citenamefont
  {{Dymnikova}}, \citenamefont {{Popov}},\ and\ \citenamefont
  {{Zentsova}}}]{1982Ap&SS..85..231D}%
  \BibitemOpen
  \bibfield  {author} {\bibinfo {author} {\bibfnamefont {I.~G.}\ \bibnamefont
  {{Dymnikova}}}, \bibinfo {author} {\bibfnamefont {A.~K.}\ \bibnamefont
  {{Popov}}},\ and\ \bibinfo {author} {\bibfnamefont {A.~S.}\ \bibnamefont
  {{Zentsova}}},\ }\bibfield  {title} {\bibinfo {title} {{Bursts of
  Gravitational Radiation from Active Galactic Nuclei and Globular Clusters}},\
  }\href {https://doi.org/10.1007/BF00653445} {\bibfield  {journal} {\bibinfo
  {journal} {Astrophysics and Space Science}\ }\textbf {\bibinfo {volume}
  {85}},\ \bibinfo {pages} {231} (\bibinfo {year} {1982})}\BibitemShut
  {NoStop}%
\bibitem [{\citenamefont {{O'Leary}}\ \emph {et~al.}(2009)\citenamefont
  {{O'Leary}}, \citenamefont {{Kocsis}},\ and\ \citenamefont
  {{Loeb}}}]{2009MNRAS.395.2127O}%
  \BibitemOpen
  \bibfield  {author} {\bibinfo {author} {\bibfnamefont {R.~M.}\ \bibnamefont
  {{O'Leary}}}, \bibinfo {author} {\bibfnamefont {B.}~\bibnamefont
  {{Kocsis}}},\ and\ \bibinfo {author} {\bibfnamefont {A.}~\bibnamefont
  {{Loeb}}},\ }\bibfield  {title} {\bibinfo {title} {{Gravitational waves from
  scattering of stellar-mass black holes in galactic nuclei}},\ }\href
  {https://doi.org/10.1111/j.1365-2966.2009.14653.x} {\bibfield  {journal}
  {\bibinfo  {journal} {Monthly Notices of the Royal Astronomical Society}\
  }\textbf {\bibinfo {volume} {395}},\ \bibinfo {pages} {2127} (\bibinfo {year}
  {2009})},\ \Eprint {https://arxiv.org/abs/0807.2638} {arXiv:0807.2638
  [astro-ph]} \BibitemShut {NoStop}%
\bibitem [{\citenamefont {{Bae}}\ \emph {et~al.}(2017)\citenamefont {{Bae}},
  \citenamefont {{Lee}}, \citenamefont {{Kang}},\ and\ \citenamefont
  {{Hansen}}}]{2017PhRvD..96h4009B}%
  \BibitemOpen
  \bibfield  {author} {\bibinfo {author} {\bibfnamefont {Y.-B.}\ \bibnamefont
  {{Bae}}}, \bibinfo {author} {\bibfnamefont {H.~M.}\ \bibnamefont {{Lee}}},
  \bibinfo {author} {\bibfnamefont {G.}~\bibnamefont {{Kang}}},\ and\ \bibinfo
  {author} {\bibfnamefont {J.}~\bibnamefont {{Hansen}}},\ }\bibfield  {title}
  {\bibinfo {title} {{Gravitational radiation driven capture in unequal mass
  black hole encounters}},\ }\href {https://doi.org/10.1103/PhysRevD.96.084009}
  {\bibfield  {journal} {\bibinfo  {journal} {\prd}\ }\textbf {\bibinfo
  {volume} {96}},\ \bibinfo {eid} {084009} (\bibinfo {year} {2017})},\ \Eprint
  {https://arxiv.org/abs/1701.01548} {arXiv:1701.01548 [gr-qc]} \BibitemShut
  {NoStop}%
\bibitem [{\citenamefont {{Leigh}}\ \emph {et~al.}(2018)\citenamefont
  {{Leigh}}, \citenamefont {{Geller}}, \citenamefont {{McKernan}},
  \citenamefont {{Ford}}, \citenamefont {{Mac Low}}, \citenamefont
  {{Bellovary}}, \citenamefont {{Haiman}}, \citenamefont {{Lyra}},
  \citenamefont {{Samsing}}, \citenamefont {{O'Dowd}}, \citenamefont
  {{Kocsis}},\ and\ \citenamefont {{Endlich}}}]{2018MNRAS.474.5672L}%
  \BibitemOpen
  \bibfield  {author} {\bibinfo {author} {\bibfnamefont {N.~W.~C.}\
  \bibnamefont {{Leigh}}}, \bibinfo {author} {\bibfnamefont {A.~M.}\
  \bibnamefont {{Geller}}}, \bibinfo {author} {\bibfnamefont {B.}~\bibnamefont
  {{McKernan}}}, \bibinfo {author} {\bibfnamefont {K.~E.~S.}\ \bibnamefont
  {{Ford}}}, \bibinfo {author} {\bibfnamefont {M.~M.}\ \bibnamefont {{Mac
  Low}}}, \bibinfo {author} {\bibfnamefont {J.}~\bibnamefont {{Bellovary}}},
  \bibinfo {author} {\bibfnamefont {Z.}~\bibnamefont {{Haiman}}}, \bibinfo
  {author} {\bibfnamefont {W.}~\bibnamefont {{Lyra}}}, \bibinfo {author}
  {\bibfnamefont {J.}~\bibnamefont {{Samsing}}}, \bibinfo {author}
  {\bibfnamefont {M.}~\bibnamefont {{O'Dowd}}}, \bibinfo {author}
  {\bibfnamefont {B.}~\bibnamefont {{Kocsis}}},\ and\ \bibinfo {author}
  {\bibfnamefont {S.}~\bibnamefont {{Endlich}}},\ }\bibfield  {title} {\bibinfo
  {title} {{On the rate of black hole binary mergers in galactic nuclei due to
  dynamical hardening}},\ }\href {https://doi.org/10.1093/mnras/stx3134}
  {\bibfield  {journal} {\bibinfo  {journal} {Monthly Notices of the Royal
  Astronomical Society}\ }\textbf {\bibinfo {volume} {474}},\ \bibinfo {pages}
  {5672} (\bibinfo {year} {2018})},\ \Eprint {https://arxiv.org/abs/1711.10494}
  {arXiv:1711.10494 [astro-ph.GA]} \BibitemShut {NoStop}%
\end{thebibliography}%

\end{document}